\documentclass{emulateapj}
\usepackage{natbib}
\usepackage{amsmath}
\usepackage[section]{placeins}
\begin{document}

\title{Fast Star, Slow Star; Old Star, Young Star: Subgiant Rotation as a Population and Stellar Physics Diagnostic}
\shorttitle{}

\author{Jennifer L. van Saders and Marc H. Pinsonneault}
\affil{Department of Astronomy, The Ohio State University, 140 West 18th Avenue, Columbus, OH, 43210}
\email{vansaders@astronomy.ohio-state.edu}

\shortauthors{van Saders \& Pinsonneault}
\slugcomment{Submitted to ApJ}

\begin{abstract}

Stellar rotation is a strong function of both mass and evolutionary state. Missions such as \textit{Kepler} and \textit{CoRoT} provide tens of thousands of rotation periods, drawn from stellar populations that contain objects at a range of masses, ages, and evolutionary states. Given a set of reasonable starting conditions and a prescription for angular momentum loss, we address the expected range of rotation periods for cool field stellar populations. We find that cool stars fall into three distinct regimes in rotation. Rapid rotators with surface periods less than 10 days are either young low-mass main sequence (MS) stars, or higher mass subgiants which leave the MS with high rotation rates. Intermediate rotators (10-40 days) can be either cool MS dwarfs, suitable for gyrochronology, or crossing subgiants at a range of masses. Gyrochronology relations must therefore be applied cautiously, since there is an abundant population of subgiant contaminants. The slowest rotators, at periods greater than 40 days, are lower mass subgiants undergoing envelope expansion. We identify additional diagnostic uses of rotation periods. There exists a period-age relation for subgiants distinct from the MS period-age relations. There is also a period-radius relation that can be used as a constraint on the stellar radius, particularly in the interesting case of planet host stars. The high-mass/low-mass break in the rotation distribution on the MS persists onto the subgiant branch, and has potential as a diagnostic of stellar mass. Finally, this set of theoretical predictions can be compared to extensive datasets to motivate improved modeling.

\end{abstract}
\keywords{stars: rotation --- stars: evolution --- stars: fundamental parameters --- stars: interiors}

\section{Introduction} \label{sec:introduction}

Stellar rotation is observed to be both a strong function of mass and evolutionary state. The rotation rates of stars respond to the loss of angular momentum over time and changes to the stellar structure as objects evolve. Furthermore, the fundamental structural differences between hot and cool stars, namely the presence or absence of a convective envelope, produce period distributions with a strong dependence on mass. The fact that rotation is such a sensitive function of stellar mass and evolutionary state means that it can be used as a powerful tool to understand the stars themselves. Rotation is already routinely exploited as a diagnostic of stellar ages in gyrochronology \citep{barnes2007,mamajek2008,meibom2009, meibom2011}, in which the strong stellar-spin down due to magnetized winds in cool MS stars is used as a clock.  Our focus is different: we aim to show that rotation is not only a useful diagnostic for these cool main sequence stars, but also a powerful tool for studying hot and evolved stars. 

On the main sequence there are two rotational regimes, distinguished by effective temperature. Cool stars ($< 6200$ K) are typically slow rotators, with periods of 10's of days. These rotation periods are far slower than those one would expect were the star to conserve its angular momentum throughout its pre-main sequence collapse and contraction, and far slower than observed rotation periods in very young clusters (compare the Mt. Wilson sample of old field stars in \citealt{baliunas1983} to \citealt{hartman2010} for the Pleiades). Furthermore, observations of the rotation period distributions in clusters at a number of different ages have confirmed that these cool stars lose angular momentum and spin down over time \citep[for an extensive compilation, see][]{irwin2009}. This angular momentum loss proceeds via magnetized stellar winds \citep{weber1967,schatzman1962} and is often parameterized using loss laws of the form $dJ/dt \propto \omega^3$ \citep{kawaler1988,krishnamurthi1997, sills2000}. The angular momentum loss scales strongly with the rotation velocity, which forces a convergence in the rotation periods for all objects of the same mass at late times: objects born with slow rotation rates lose angular momentum slowly, whereas those born rapidly rotating quickly spin down. Therefore, even a population of cool stars born with a wide range of rotation periods will be slowly rotating with a narrow range of allowed periods at late times \citep{epstein2012}. It is this characteristic of the spin-down that makes gyrochronology possible: it ensures both that rotation will be a strong function of age, and that sensitivity to initial conditions for old objects will be minimal. 

In contrast, hot stars ($>6200$K) are rapidly rotating. This transition between slow and rapid rotation, known as the Kraft break \citep{kraft1967}, occurs at roughly $\sim 1.3 \textrm{M}_{\odot}$, where surface convective envelopes become vanishingly thin, and the stars are unable to generate the magnetic winds and drive angular momentum loss. As a result, these objects are rapidly rotating from birth with periods that evolve only mildly over the course of the main sequence. Studies that observed stars at or beyond this boundary have confirmed this general picture \citep{melo2001,wolff1997,zorec2012}. 

While the evolution of the rotation periods on the main sequence is driven by the presence (or lack of) angular momentum loss through winds, structural changes become an important contributor for evolved stars. As stars leave the main sequence their cores contract and envelopes expand, resulting in an increase in their moments of inertia. Even in the most simplistic case in which the star conserves angular momentum and rotates rigidly, this increase in the moment of inertia would lead to a decrease in the surface rotation rate as the star evolves across the subgiant branch. The addition of winds on the subgiant branch allow for further spin-down. Because objects above the Kraft break develop convective envelopes on the subgiant branch, there can be wind-driven angular momentum loss for both main sequence rotational regimes in evolved stages. 

In the era of \textit{Kepler} \citep{borucki2010} and \textit{CoRoT} \citep{baglin2006} we will have access to thousands to tens of thousands of rotation periods \citep[][Garc\'{i}a et al. 2013, in prep.]{affer2012,mcquillan2013,nielsen2013}, all of which will be drawn from a population of stars with many different masses, ages, and evolutionary states. In order to understand the underlying stellar populations and interpret their rotation distributions, we first need a basic set of theoretical predictions for the behavior of rotation across large regions of the HR diagram. Our approach is to choose a prescription for tracking rotational evolution that minimizes our model complexity, allowing us to produce basic predictions for the rotation periods of models across a wide range of masses with minimal assumptions. We will assume that all stars rotate as solid bodies at all times, and that angular momentum is lost through magnetized winds for all objects with sufficiently thick convective envelopes. We draw our initial distribution of rotation periods (as a function of mass) from open clusters where the sensitivity to the somewhat complex pre-main sequence evolution is minimal.

With this set of predictions we show that the strong mass dependence of the surface rotation rate on the main sequence persists onto the subgiant branch, and provides a useful test for the mass scale in asteroseismology and means of identifying more massive stars in the crossing region. By considering evolved objects we show both that there is a strong period-age relationship among subgiant stars that has not been appreciated in the literature, and that the application of the standard gyrochronology relations must be done with some care to avoid mistaking a crossing subgiant for an unevolved dwarf and applying an inappropriate period-age relationship. Additionally, the importance of the envelope expansion in determining the rotation period of stars on the subgiant branch results in a strong correlation between rotation period and radius, which will be particularly useful in the case of transiting planet hosts, in which errors on the stellar radius translate directly into errors on the planet size. Finally, once we are able to compare large datasets of rotation periods to these model predictions, disagreements between our models and the data will motivate the inclusion of missing physics and additional model complexity.  

In Section \ref{sec:methods} we describe the construction of our model grid and treatment of angular momentum evolution, as well as grids to test the effects of variations on the standard physics. In Section \ref{sec:results} we show that our simple assumptions preserve the observed rotational properties of both cool and hot stars, and in particular that the Kraft break is naturally recovered. We discuss the manner in which reasonable variations in the input physics affect our results. In Section \ref{sec:tool} we elucidate the many ways in which we can use stellar rotation as a tool in the context of stellar populations: as a test of the mass scale, age indicator, and constraint on stellar radii. The relationship of these new tools to the existing framework of gyrochronology is discussed. We discuss uncertainties in our models and future prospects in Section \ref{sec:discussion}. Finally, we conclude in Section \ref{sec:conclusion}.

\section{Methods} \label{sec:methods}

In order to investigate the evolution of the surface rotation rates of a population of stars over time we must specify a reasonable set of starting conditions, here empirically motivated by data from open clusters. We also need a prescription by which we evolve these initial conditions forward in time, which includes processes such as angular momentum (AM) loss, structural evolution, and internal AM transport as a function of mass and composition. In the following section we motivate our initial conditions and prescription for AM evolution. We use observed rotation periods in clusters to calibrate our AM loss law to reproduce both the upper and lower envelopes of the rotation distribution. This calibration sets our starting conditions, and we apply this loss law to a grid of models to examine rotation as a function of age and mass.

\subsection{Simplifying Assumptions}

Treatments of AM evolution can be complex. We opt to construct a set of models based on the simplest set of assumptions that are consistent with the data. There are strong trends in the rotation distribution as a function of effective temperature and gravity that should exist for a variety of reasonable AM evolution models. We choose the simplest case so that we can first investigate the expected range of rotation rates in a base case. 

Stars are born with a range of initial rotation periods, and pre-MS AM evolution is complex. At ages less than $\sim 5-10$ Myr stars are thought to shed angular momentum through coupling with their protostellar disks. Both accretion-powered stellar winds \citep{matt2005} and disk-locking \citep{koenigl1991, shu1994} are possible mechanisms for the coupling, but in either case the stellar AM is regulated by interaction with the disk on the timescale of the disk lifetime. The duration of this disk-locking period differs among stars, resulting in a spread in rotation rates, with rapid rotators having been released from their disks at earlier times than slow rotators. There is also evidence that the AM loss law saturates for rapid rotators \citep{sills2000, krishnamurthi1997}, and that differential rotation with depth is necessary to reproduce the young cluster data \citep{denissenkov2010}.

Rotational evolution on the MS is comparatively simple. Cool stars below the Kraft break rotate slowly, and have therefore forgotten their initial conditions by $\sim0.5$ Gyr \citep{pinsonneault1989}. Hot stars above the break do not lose AM and the scatter in initial rotation period persists. We have little information about the internal rotation profiles of stars on the MS save that of the Sun. However, the Sun \citep[down to $0.2 R_{\odot}$][]{thompson2003} and the rapidly rotating upper envelope of open cluster stars are well approximated as a solid bodies, although uncertainties about the validity of the solid body treat remain for evolved and hot stars \citep[see][]{zorec2012,deheuvels2012}. In cases in which there are strong internal magnetic fields, solid body rotation is the expectation \citep{spruit1987}.

We therefore make the following simplifying assumptions. We draw our starting conditions from intermediate age open clusters ($\sim 0.6$ Gyr) M37, Praesepe, and the Hyades, supplemented with data from the Pleiades (125 Myr) for rapid rotators. The upper and lower bounds of this period distribution give us a fair indicator of the mean and range of periods as a function of mass, which accounts for the complex pre-MS evolution without requiring that we model it. This distribution can be evolved forward under the simplifying assumption of solid body rotation on the MS for ages $> 0.5$ Gyr. We carry the assumption of solid body rotation onto the SGB in our base model.    

We also make several simplifying assumptions in the stellar modeling itself. We do not construct fully rotating models. Stellar parameters are tracked with underlying non-rotating models that do not include diffusion, overshoot, or the effects of rotational mixing or deformation. The evolution of the surface rotation period is tracked separately assuming a starting period and appropriate AM loss law. In the following sections we detail our methods for extracting starting conditions from cluster period distributions, the form or our AM loss law on the MS and SGB, and the underlying stellar models that we use to track stellar parameters. 

\subsection{Starting Conditions} \label{sec:initial_conditions}

In practice, to compile a set of rotation periods across a large mass range ($0.4-2.0M_{\odot}$) we must look to several datasets collected in fundamentally different ways. Cool stars are heavily spotted, and therefore have significant spot-modulation in their light curves. Dedicated photometric surveys are capable of determining rotation periods through regular observations of a set of stars. Hot stars, however, generally have only a few spots, if any, and rotation rates for these stars are instead determined using rotational broadening of spectral lines, which are subject to a $\sin{i}$ ambiguity. We are therefore forced to draw data from multiple sources to cover the entire mass range of interest. 

Open clusters represent a rich source of empirical data for our purposes. Many are well-studied, with well-determined ages and rotation rates or periods for hundreds of stars. For our exercise we choose M37 \citep[550 Myr, metallicity of +0.045,][]{hartman2009} and the Pleiades \citep[125 Myr, metallicity of +0.003][]{hartman2010}. For M37 we utilize the rotation periods from \citet{hartman2009} obtained from the spot modulation of stellar light curves. These observations consist of the 375 objects in the ``clean'' sample for which we determine stellar masses using the cluster parameters from \citep{hartman2008} and colors using the \citet{an2007} isochrones. Objects fall in the mass range of $0.4 < \textrm{M/M}_{\odot} < 1.2$. The Pleiades sample is drawn from \citep{hartman2010}, also from spot-modulation periods. We use the measured rotation periods and inferred masses for all non-binary members in the \citeauthor{hartman2010} sample without adjustment. 

A hot star sample is drawn from the Hyades and Praesepe clusters, both of which are similar in age to M37 \citep{an2007}. When we consider the hot stars, the small age differences between these clusters are not important for rotation. We use the WEDBA\footnotemark{} open cluster database to draw $v\sin{i}$'s for Hyades members from \citet{kraft1965}, \citet{abt1995}, \citet{reid2000}, and \citet{paulson2003} and cluster membership from \citet{perryman1998}. Any objects noted as spectroscopic binaries in \citet{perryman1998} were omitted from the sample. Average values of $V$ and $B-V$ for each star were drawn from the compilation of \citet{mermilliod1995}, and we select only objects with $B-V < 0.8$. In total, 76 objects have $v \sin{i}$'s, membership information, and photometry from the named sources, and 65 of those objects have $B-V < 0.8$. We draw $v\sin{i}$'s for Praesepe from \citet{treanor1960} and \citet{mcgee1967} and photometry from \citet{johnson1952} and \citet{dickens1968}. Object flagged as binaries in the WEBDA database are omitted. A total of 54 objects are selected, 36 of which have $B-V < 0.8$. \footnotetext{http://www.univie.ac.at/webda/}.

\begin{deluxetable} {llllll}
\tabletypesize{\scriptsize}
\tablecolumns{5}
\tablewidth{0pt}
\tablecaption{M37 rotation data}
\tablehead{
                        \colhead{Object \tablenotemark{a}}     &
                        \colhead{$V$(mag)}      &
		        \colhead{$B-V$(mag)}		&
			\colhead{M/$\textrm{M}_{\odot}$ \tablenotemark{b}}		&
			\colhead{P(days)} \tablenotemark{c} }
\startdata
2  &  17.735  &  1.144  &  0.864  &  8.360  \\ 
4  &  19.645  &  1.480  &  0.711  &  1.701  \\ 
5  &  17.325  &  1.022  &  0.942  &  7.775  \\ 
6  &  17.099  &  0.990  &  0.964  &  16.715  \\ 
8  &  17.877  &  1.165  &  0.853  &  8.474  \\ 
\enddata
\tablenotetext{a}{Object designation in the original \citet{hartman2009} tables.}
\tablenotetext{b}{Mass inferred from the combination of $V$ and $B-V$ using the \citet{an2007} isochrones assuming the cluster parameters: $t = 550$ Myr, $\textrm{[Fe/H]} = 0.045$, $E(B-V) = 0.227$, and distance modulus $\mu = 11.57$}
\tablenotetext{c}{periods from the AoV algorithm in \citet{hartman2009} with $N = 2$}
\tablecomments{This table is available in its entirety in a machine-readable form in the online journal. A portion is shown here for guidance regarding its form and content.}
\label{tbl1}
\end{deluxetable}

\begin{deluxetable*} {lllllll}
\tabletypesize{\scriptsize}
\tablecolumns{7}
\tablewidth{0pt}
\tablecaption{Hyades \& Praesepe rotation data}
\tablehead{
                        \colhead{Object \tablenotemark{a}}     &
                        \colhead{$B-V$(mag)\tablenotemark{b}}      &
		        \colhead{$\textrm{M/M}_{\odot}$\tablenotemark{c}}		&
			\colhead{$\textrm{R/R}_{\odot}$\tablenotemark{c}}		&
			\colhead{$v\sin{i}$(km/s)} \tablenotemark{b}  &
			\colhead{Period (days)} \tablenotemark{d}  &
			\colhead{Source \tablenotemark{e}} }
\startdata
1  &  0.570  &  1.199  &  1.147  &  5.5  &  10.557  &  1  \\ 
2  &  0.620  &  1.140  &  1.066  &  5.5  &  9.807  &  1  \\ 
6  &  0.339  &  1.584  &  1.639  &  56.0  &  1.481  &  1  \\ 
8  &  0.412  &  1.452  &  1.497  &  50.0  &  1.515  &  1  \\ 
10  &  0.597  &  1.167  &  1.101  &  6.2  &  8.991  &  1  \\ 
\enddata
\tablenotetext{a}{Object designation in WEBDA for each cluster, respectively.}
\tablenotetext{b}{Averages over values available in \citet{mermilliod1995}.}
\tablenotetext{c}{Mass and radius inferred using $B-V$ and the \citet{an2007} isochrones assuming the cluster parameters: $t = 550(550)$ Myr, $\textrm{[Fe/H]} = 0.13(0.11)$, $E(B-V) = 0.003(0.007)$ for the Hyades(Praesepe).}
\tablenotetext{d}{under that assumption that $\sin{i} =1$, and using the inferred stellar radius.}
\tablenotetext{e}{1:Hyades, 2: Praesepe}
\tablecomments{This table is available in its entirety in a machine-readable form in the online journal. A portion is shown here for guidance regarding its form and content.}
\label{tbl2}
\end{deluxetable*}

\subsection{Structural Evolution}
We track the evolution of the moment of inertia with stellar models created using the Yale Rotating Evolution Code \citep[YREC, see][]{vansaders2012,pinsonneault1989, bahcall1992, bahcall1995, bahcall2001} for $0.4 < M/M_{\odot} < 2.0$. Models are evolved onto the RGB or to 30 Gyr, whichever occurs first. The models do not include helium or heavy element diffusion. We use the the atmosphere and boundary conditions of \citet{kurucz1997} \footnotemark, nuclear reaction rates of \citet{adelberger2011} with weak screening \citep{salpeter1954}, and employ the mixing length theory of convection \citep{cox1968, vitense1953} with neither core nor envelope overshoot (primarily important for stars with $M > 1.2 M_{\odot}$). Opacities are from the Opacity Project (OP) \citep{mendoza2007} for a \citet{grevesse1998} solar mixture, supplemented with the low temperature opacities of \citet{ferguson2005}. We utilize the 2006 OPAL equation of state \citep{rogers1996,rogers2002}. 

We use a solar calibration to set the value of the mixing-length parameter, $\alpha$, and the initial composition, $X$, $Y$, and $Z$ such that a $1.0 M_{\odot}$ model at 4.57 Gyr \citep[see][]{bahcall1995} recovers the solar radius, luminosity, and surface abundance of $R_{\odot}=6.9598\times10^{10} \textrm{ cm}$, $L_{\odot}=3.8418\times10^{33} \textrm{ ergs}\textrm{ s}^{-1}$,  and $Z/X = 0.02289 $ from \citet{grevesse1998}. A calibration using this standard set of physics yields $\alpha =1.8098 $, $X = 0.71943$, and $Z = 0.01646$. We define the end of the MS as the age at which the core hydrogen fraction drops below $X_c = 0.0002$, and the end of the subgiant branch is defined mathematically as the local minimum on a $\log{g}-T_{eff}$ diagram in the evolutionary tracks at the base of the giant branch. These stellar models provide the necessary information about radius, luminosity, and moment of inertia as a function of mass and age, but do not include the structural effects of rotational deformation, or account for the effects of rotational mixing. 

\footnotetext{In models whose effective temperature exceeds the maximum temperature of the Kurucz tables, $\log{T} = 3.95$, the code switches the atmosphere to grey (at $\sim1.89 M_{\odot}$ for $\textrm{[Fe/H]} = 0.0$ and $\sim1.74 M_{\odot}$ for $\textrm{[Fe/H]} = -0.2$). The differences between the grey and Kurucz atmospheres are not important for the hot stars, but induce shifts in the location of the giant branch on the HR diagram. We therefore run these models are run in two stages: they are evolved from PMS until they cross the $\log{T} = 3.95$ boundary with a Kurucz/grey atmosphere. The models (now with $\log{T_{eff}}< 3.95$ are then restarted with Kurucz atmospheres and run across the SGB and onto the giant branch. This minimizes, but does not eliminate, the shift in the location of the giant branch induced by the switch between model atmospheres.} 

\subsection{Angular Momentum Transport and Loss} \label{sec:AMloss}
We must define a set of rules by which we evolve our starting conditions forward in time, which includes both assumptions about the internal transport of AM and the loss of AM via magnetized winds. We will make the assumption of efficient transport of AM for all masses on the MS and SGB; this amounts to treating all models as rigid rotators (in both radiative and convective zones). We assume that there is angular momentum loss whenever the star supports a sufficiently thick convective envelope. In practice this means that cool stars are subject to magnetic braking throughout their MS lifetimes and onto the SGB. Hot stars born above the Kraft break have very thin convective envelopes on the MS and therefore do not spin down on the MS, but do develop winds on the SGB as their convective envelopes deepen. 

There are a number of prescriptions for angular momentum loss through magnetized winds in the literature. \citet{kawaler1988} assumes that wind losses are proportional to the magnetic flux, and arrives at the functional form $dJ/dt \propto \omega^3$. Later modifications to this law, such as those in \citet{krishnamurthi1997} and \citet{sills2000}, allow for a Rossby scaling and saturation for rotation rates above some $\omega_{crit}$, both of which produce better agreement with data from open clusters. In comparison, the \citet{reiners2012} loss law assumes that the loss is proportional to the magnetic field strength, and therefore derives a form of the loss law that is far more radius-dependent than Kawaler-type variants. However, both formulations fail to adequately reproduce the rapid rotation of stars near the cool side of the Kraft break. 

We adopt the \citet{matt2008} magnetized wind formulation as updated in \citet{matt2012}. We scale the coronal magnetic field strength and mass loss rate relative to the Sun following the prescription of Pinsonneault et al. 2013 (in prep., hereafter PMM). To briefly summarize, the magnetic field is assumed to be linearly proportional to the rotation rate up to a saturation threshold $\omega_{crit}$, and proportional to the square root of the atmospheric pressure (due to equipartition considerations). The mass loss rate is assumed proportional to $L_x$ \citep{wood2005}, and we adopt $L_x/L_{bol} \sim \omega^2$ \citep{pizzolato2003}, again up to a saturation threshold.  We use a Rossby-scaled saturation threshold normalized at the Sun. We adopt a $dJ/dt$ of the form:

\begin{equation}
 \displaystyle \frac{dJ}{dt} = \left\{
	\begin{array}{l l}
	 \displaystyle f_K K_M \omega \left(\frac{\omega_{crit}}{\omega_{\odot}}\right)^2 \quad \omega_{crit} \leq \omega \frac{\tau_{cz}}{\tau_{cz, \odot}}\\
	 \displaystyle f_K K_M \omega \left(\frac{\omega \tau_{cz}}{\omega_{\odot} \tau_{cz, \odot}}\right)^2 \quad \omega_{crit} > \omega \frac{\tau_{cz}}{\tau_{cz, \odot}}\\
	\end{array} \right. 
	\label{eqn:wind_law}
\end{equation}

with

\begin{equation}
  \begin{split}
 \displaystyle \frac{K_M}{K_{M,\odot}} = c(\omega) \left(\frac{R}{R_{\odot}}\right)^{3.1} \left(\frac{M}{M_{\odot}}\right)^{-0.22} \left(\frac{L}{L_{\odot}}\right)^{0.56} \\
 \times \left(\frac{P_{phot}}{P_{phot,\odot}}\right)^{0.44}.
 \end{split}
\end{equation}

$f_K$ is a constant factor used to scale the loss law to reproduce solar the solar rotation period at solar age, $\omega_{crit}$ is the saturation threshold, $\tau_{cz}$ is the convective overturn timescale, and $P_{phot}$ the pressure at the photosphere. The term $c(\omega)$ is drawn from the \citet{matt2012} prescription and sets the centrifugal correction; it is a small correction for slowly rotating stars, and we set $c(\omega) = const = 1$. 

In practice we track the surface convective envelopes of our models only to a limiting depth, which becomes a problem for stars near the Kraft break. The model fitting point between the envelope and interior cannot be set at arbitrarily small masses (and thus depths) without inducing numerical instability, which in practice means that convective envelopes are tracked only until they pass beyond the fitting point. We have chosen a default fitting point of $\log(M_{interior}/ M_{total}) = -10^{-8}$ for all of our models, and assume that AM loss ceases when the base of the convection zone moves beyond the fitting point. This value of the fitting point was chosen such that if we assume instead that $\tau_{cz}$ is held constant (as opposed to set to zero) at its last value before the CZ moved beyond the fitting point, that the difference between the constant and zero $\tau_{cz}$ model periods would be less than one day in the periods on the MS and SGB. Note that this is unimportant for low mass stars, which always have thick convective envelopes, and for high mass stars far above the Kraft break, where convective zones are always negligibly thin on the MS. 

\subsubsection{Calibration of the Loss Law} \label{sec:methods_calibration} 
We must tune the parameters in our loss law to produce reasonable torques and to reproduce cluster and solar data. We consider two fits to the data: a fit to the rapidly rotating envelope, and a fit to the mostly slowly rotating stars, in order to characterize the range of reasonable rotation periods as a function of mass and time. The fits include 4 free parameters: the saturation threshold $\omega_{crit}$, the scaling factor $f_K$, used to normalize the loss rate to reproduce the solar value, the initial disk period $P_{disk}$, and disk-locking timescale $\tau_{disk}$. We use the Pleiades at 125 Myrs, M37 at 550 Myrs and the Sun with a rotation rate of $P_{\odot} = 25.4$ days at $t_{\odot} = 4.57$ Gyr as the reference datasets for the rapid rotators.

We fit the most rapidly rotating sequence allowing all 4 free parameters to vary. We determine the mean rotation rates for the most rapidly rotating 10\% of stars in each cluster in bins of 0.1 $\textrm{M}_{\odot}$ in the mass range of $0.45 \leq \textrm{M/M}_{\odot} \leq 1.15$, with uncertainties on the mean determined using 1000 bootstrap resamplings (shown as large, red points in Fig. \ref{fig:clusters}). We then fit the loss law models to the combined sample of these binned data points in the Pleiades, M37, and the single point for the Sun. The fit amounts to finding a combination of $\omega_{crit}$, $f_K$, $P_{disk}$, and $\tau_{disk}$ that simultaneously reproduces the rapidly rotating sequences for all masses, at all times (125 Myr, 550 Myr, and 4.57 Gyr). The fit the optimization is performed using the non-linear least squares IDL fitting package MPFIT \citep{markwardt2009}. We infer the best-fit constants $f_K =6.575$ and $\omega_{crit} = 3.394 \times 10^{-5}$ for a disk-locking timescale of $\tau_{disk}=0.2810$ Myr and period of $P_{disk} =8.134 $ days for the rapidly rotating sequence.

For the slowly rotating sequence we allow only $P_{disk}$ and $\tau_{disk}$ to vary, exclude the Pleiades from the fit, and limit the fit to mass above $0.8 M_{\odot}$ since slow rotators in this young cluster are affected by core-envelope decoupling, which is not accounted for in our models, and is more important for low masses \citep{denissenkov2010}. We add additional points to the fit at masses above the Kraft break ($1.5-1.9 M_{\odot}$ in steps of $0.1 M_{\odot}$) with rotation velocities of 50 km/s to capture the behavior of the slowly rotating A-star sequence \citep{abt1995}. We infer a disk-locking timescale of $\tau_{disk}=5.425$ Myr and period of $P_{disk} =13.809 $ days for the slow rotators, while $f_K$ and $\omega_{crit}$ are held at the values derived for the fast rotators.

For comparison, we also fit a more standard modified Kawaler law \citep{krishnamurthi1997}:
\begin{equation}
 \displaystyle \frac{dJ}{dt} = \left\{
	\begin{array}{l l}
	 \displaystyle f_K K_w \omega_{crit}^2 \omega \left(\frac{R}{R_{\odot}}\right)^{\frac{1}{2}} \left(\frac{M}{M_{\odot}}\right)^{- \frac{1}{2}} \quad \omega > \omega_{crit} \\
	 \displaystyle f_K K_w \omega^3  \left(\frac{R}{R_{\odot}}\right)^{\frac{1}{2}} \left(\frac{M}{M_{\odot}}\right)^{-\frac{1}{2}} \quad \omega \leq \omega_{crit} \\
	\end{array} \right. 
	\label{eqn:kawaler}
\end{equation}
using the same procedure, with the same free parameters. For the modified Kawaler law we infer $f_K = 2.656$ and $\omega_{crit} = 3.126 \times 10^{-5}$, which best fits the rapidly rotating sequence with a disk locking timescale of $\tau_{disk}= 0.311$ Myr and initial disk rotation period of $P_{disk} = 7.178$ days.

\subsection{Parameter Variations}

We have made a number of simplifying assumptions in our models, the importance of which should be tested. We present the mechanical details of the models we generate to address these questions in this section; the parameter variations themselves are discussed in detail in Section \ref{subsec:parameter_variations}. There are two subsets of models:

\begin{enumerate}
\item{We create an alternate set of models that do not lose angular momentum on the SGB. These models are allowed to lose AM on the MS, but conserve AM once they have passed the turnoff, defined here to be when the central hydrogen $X_c < 0.0002$.}
\item{We construct two grids of interiors models with different compositions: one at solar metallcity, and the other at the mean metallicity of the \textit{Kepler} stars, $\textrm{[Fe/H]} = -0.2$, $X = 0.73283 $, $Z = 0.01058 $. The rotational evolution of both grids is treated in exactly the same manner: the same initial conditions are applied to both populations, and the same prescription for rotational evolution is applied. We do not account for differences in the initial rotation rates of stars as a function of metallicity, which remain observationally ambiguous.} 
\end{enumerate}

\section{Results} \label{sec:results}
We present the results of our stellar models with rotation, and show that the chosen form of the wind law naturally reproduces the Kraft break, without fine-tuning. We compare our model predictions to existing observational data both in clusters and the field. Finally, in Section \ref{subsec:isochrones}, we present tables with our rotation predictions. 

\subsection{Rotation rate as a function of mass and evolutionary state} 

\subsubsection{Behavior of Representative Models}
\begin{figure}
	\centerline{\includegraphics[scale = 1.0]{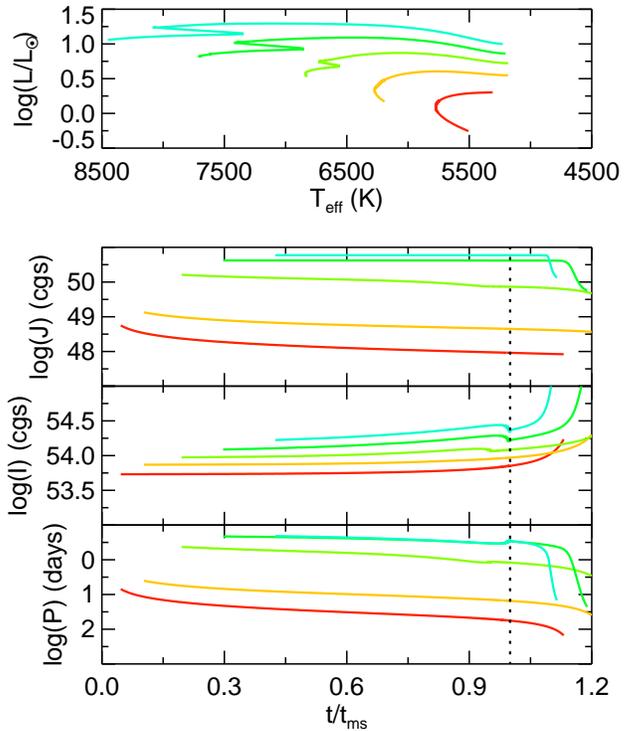}}
        \caption{Top panel: Evolutionary tracks for 1.7, 1.5, 1.3, 1.1, and 0.9 solar mass objects, with the bluest objects being the most massive. $2^{nd}$ panel: Total angular momentum as a function of MS lifetime for the same set of masses. The dotted line marks the location of the main sequence turnoff. $3^{rd}$ panel: Total moment of inertia for each of the models. Bottom panel: surface rotation period as a function of time. All models are plotted at ages $t > 0.55$ Gyr for $\textrm{[Fe/H]} = -0.2$ with the rapidly rotating launch conditions.}
	\label{fig:intro_panels}
\end{figure}

We begin by considering a small subset of models, and show predictions for the total angular momentum, moments of inertia, and rotation periods as a function of time for a few representative masses in Fig. \ref{fig:intro_panels} under the ``fast launch'' conditions. In the second panel (angular momentum as a function of time), we see that more massive stars (bluer/ greener curves) lose very little angular momentum on the main sequence, while low mass stars (red curves) undergo substantial loss. High mass stars, which have no winds on the MS, develop deep surface convection zones for the first time on the SGB and begin to lose angular momentum, marked by the steep downturn in the curves of the most massive models after the main sequence turnoff. The second important ingredient in the determination of the surface rotation rate, the moment of inertia, gradually increases over the course of the MS, followed by a steep increase on the SGB as stars undergo envelope expansion. For solid body models, an increase in the moment of inertia results directly in an increase of rotation period. The bottom panel of Figure \ref{fig:intro_panels} shows the results for these two effects combined: low mass objects spin down and are slowly rotating on the MS, and spin down further on the SGB. High mass objects do not lose angular momentum to stellar winds on the MS, are born rapidly rotating and remain so until they undergo expansion and wind losses on the SGB. 

\subsubsection{Ensemble Behavior}
 
\begin{figure*}
	\centerline{\includegraphics[scale = 0.6]{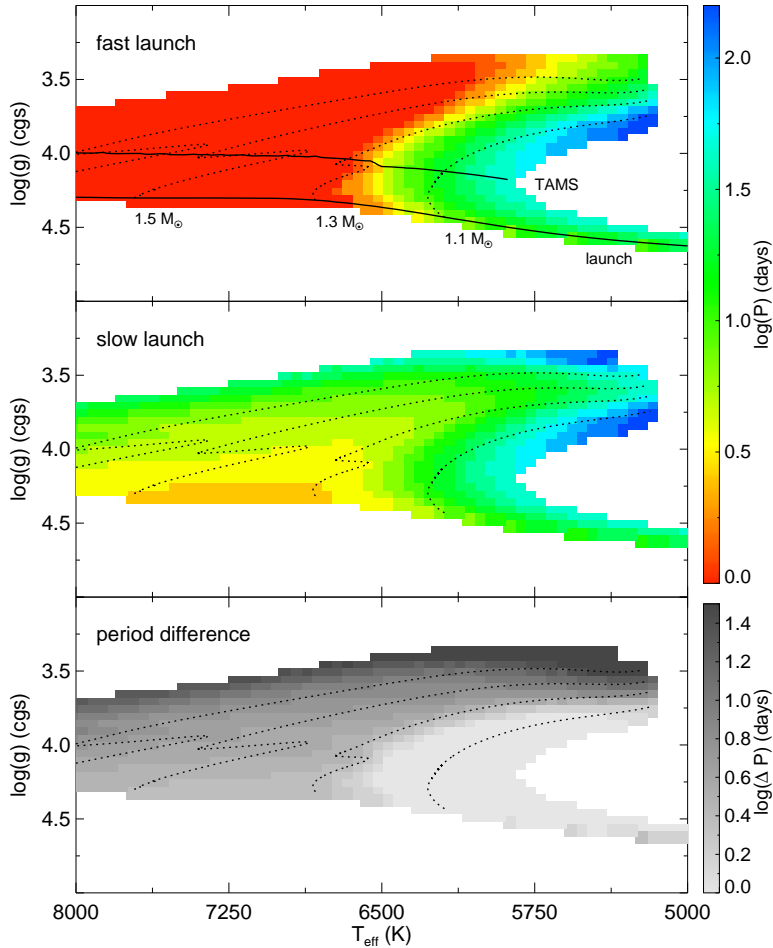}}
        \caption{Our theoretical predictions on a $\textrm{T}_{eff}$ vs. log(g) diagram. The top panel represent models launched with the rapidly rotating starting conditions, the middle the slow starting conditions, and the bottom the difference (at fixed log(g) and $T_{eff}$) between them. Color encodes the rotation period, with bluer colors representing longer rotation periods. Evolutionary tracks for benchmark masses are over plotted in black dotted lines. Solid black lines in the top panel show the launch conditions at 0.55 Gyr (``launch'') and terminal age main sequence (``TAMS''). Models are plotted for ages $0.55 < t < 10.0$ Gyr at $\textrm[Fe/H] = -0.2$.}
	\label{fig:teffvlogg_colors}
\end{figure*}

Our primary results for the full set of models are given in Figure \ref{fig:teffvlogg_colors}, where we show period (encoded by color) across many different masses and evolutionary states on a $\log{g}-T_{eff}$ diagram. The diagram is populated by selecting tiles with $ \Delta \log{g} = 0.05 $ and $\Delta T_{eff} = 50$ K, calculating the mean rotation period of all models that pass through each box in log(g)-$T_{eff}$, and coloring each box accordingly. The top panel shows the distribution for models evolved with fast launch conditions defined by the loss law fitted to the rapidly rotating cluster sequences. The middle panel shows the same, but for models evolved under the slow launch conditions. The bottom panel shows the simple difference in mean period, tile by tile, between the fast and slow starting conditions.

The location of the Kraft break is clearly visible in the top panel as the abrupt transition between red (rapidly rotating) and green (intermediate rotation) regions. One should note that the break itself is imprinted on the MS starting conditions (denoted by the curve labeled ``launch''), but persists well onto the subgiant branch, due to the differences in the angular momentum loss between stars with and without convective envelopes. In the slow launch case the same basic patterns are still present, but the contrast between the rotation rates of objects above and below the Kraft break on the MS is smaller. This contrast is further diminished by the envelope expansion and wind losses on the SGB.

The bottom panel illustrates an important feature: stars born below the Kraft break, regardless of whether they are born rapidly or slowly rotating, have a very narrow range of predicted periods on the majority of the MS and onto the SGB. This is the result of strong $dJ/dt \propto \omega^3$ dependence in the loss law: rapidly rotating objects are quickly spun down, whereas slowly rotating objects undergo less angular momentum loss. This forces a convergence in the rotation rates at relatively early times ($\sim 1$ Gyr), and renders the starting conditions unimportant. This insensitivity to initial conditions is one of the reasons that these cool dwarfs are suitable for gyrochronology. 

In contrast, the period differences between fast and slow launch conditions for objects born above the Kraft break can be substantial. These objects undergo little or no angular momentum loss, and so MS scatter in the initial rotation periods is preserved well onto the SGB.  In practice, this means that we expect to see a range of rotation periods for the hot stars, in comparison to the well-converged cool star sequence. 

\begin{figure*}
	\centerline{\includegraphics[scale = 1.0]{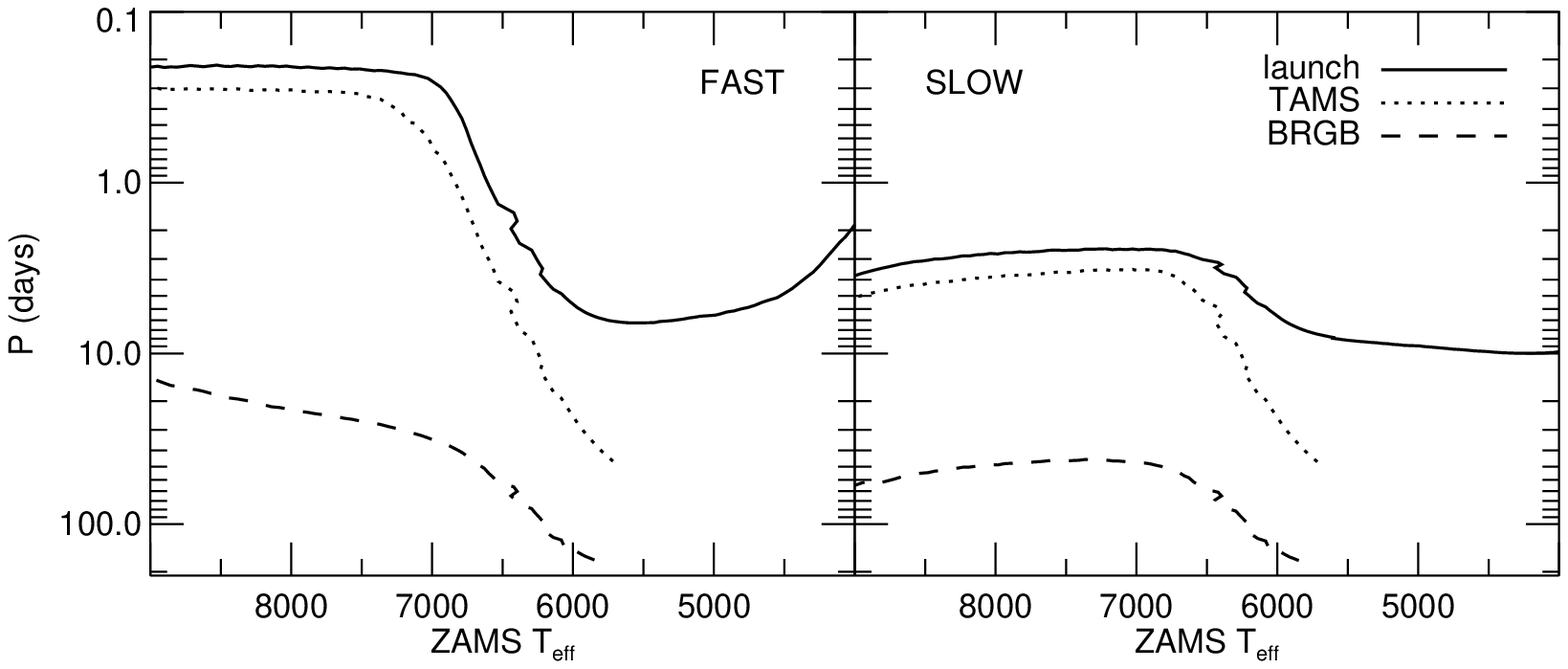}}
        \caption{Rotation period as a function of ZAMS $\textrm{T}_{eff}$ at particular evolutionary states: initial rotation periods at 550 Myr (solid line), rotation period on the TAMS (dotted line),and period at the base of the RGB (dashed line) for ages $0.55 < t < 10.0 $ Gyr.}
	\label{fig:zamsteffplot}
\end{figure*}

Figure \ref{fig:zamsteffplot} presents the same models in slices through the data at particular evolutionary stages. We show period as a function of the ZAMS $T_{eff}$ for the starting conditions, at the terminal age MS (TAMS) and end of the subgiant branch (base of the red giant branch, BRGB). In this case, all models are at the same metallicity, and so $T_{eff}$ can also be viewed as a proxy for mass. Again we see that objects above the Kraft break begin rapidly rotating, with periods of less than a day for the fast launch, compared to the cool stars which have more modest few-to-ten day periods. At the TAMS (dotted curve in Fig. \ref{fig:zamsteffplot}), the objects above the Kraft break have spun down very little, while the cool stars have periods of $10 < P < 40$ days. All objects undergo significant spin down on the SGB. The rotation period increases by two orders of magnitude for the hot stars by the BRGB (dashed curve) due the combined effects of envelope expansion and stellar winds, while cool stars undergo a similar but more modest response. The difference in the magnitude of the SGB spin down between hot and cool stars is due to two effects. First, the moment of inertia increase is far more substantial for the hot stars (see Fig. \ref{fig:intro_panels}), and second, the hot stars develop convective envelopes while rapidly rotating (in comparison to the cool stars, which always have convective envelopes, and are already rotating very slowly by the SGB), which means that the onset of stellar winds is more important for the massive stars.

\subsection{Comparison of Model Predictions with Observations}
 
\subsubsection{Agreement with Young Clusters}

\begin{figure}
	\centerline{\includegraphics[scale = 1.0]{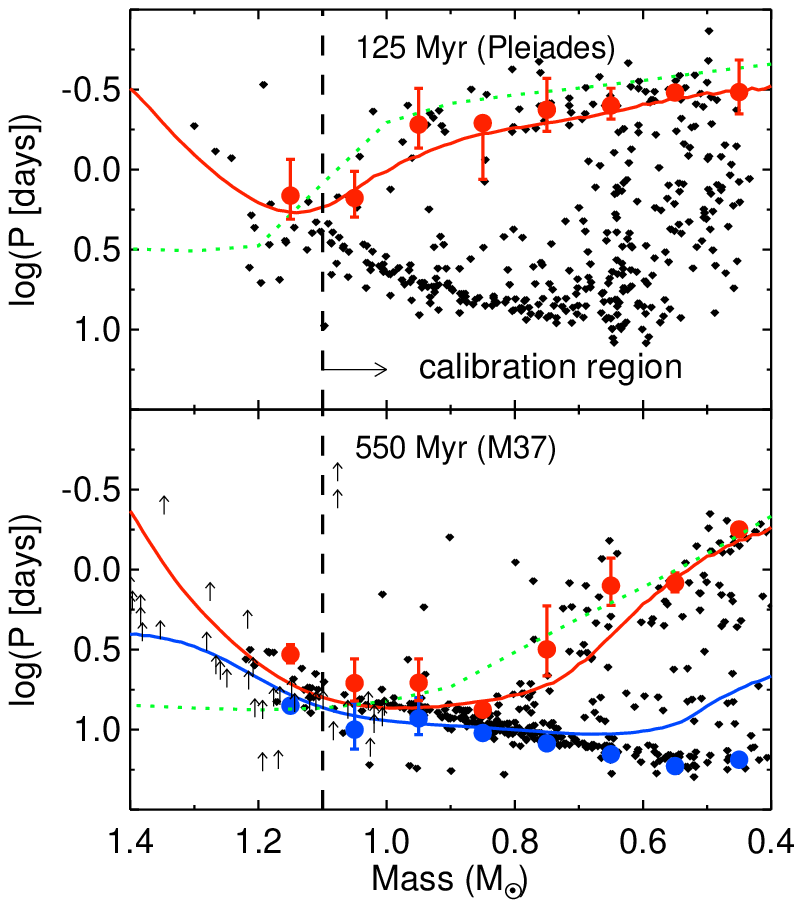}}
        \caption{Calibrated wind loss law over plotted on rotation data from the Pleiades (125 Myr) and M37 (550 Myr) (points), and $v \sin{i}$ data for the Hyades and Praesepe (not included in the fit, and plotted with a $4/ \pi$ correction factor to statistically account for a range of inclinations, plotted as arrows.). Solid blue curves show the Pinsonneault et al.(2013) loss law fit to the slowly rotating sequence at at mass intervals denoted by large blue points (10th percentile averages with bootstrapped uncertainties). Red curves represent the loss law fit to the rapidly rotating sequence, and large red points the values to which the model was fit (90th percentile averages with bootstrapped uncertainties). The dotted green curve is for a similar fit using the modified Kawaler law, fit to the rapidly rotating sequence (red points).}
	\label{fig:clusters}
\end{figure}

We compare our models to representative young and intermediate-aged open clusters in Figure \ref{fig:clusters}.  For reference we also indicate the mass dependence that would have been predicted from an extrapolation of a Rossby-scaled Kawaler-style angular momentum loss law.

The upper envelope, given by the red curve in Figure \ref{fig:clusters}, is a fit to the $0.5 < M/M_{\odot} < 1.1$ range, and is, by construction, designed to reproduce the low-mass rapidly rotating sequences in the Pleiades and M37. Objects more massive than $1.2 M_{\odot}$, which here are a mixture of Pleiades, M37, Praesepe, and Hyades stars, were not included in this fit, and yet the model agrees with the data well in this region. The PMM loss law reproduces rotation rates of massive stars far better than the Rossby-scaled Kawaler law, which predicts periods far longer than are actually observed. We naturally reproduce the Kraft break with parameters derived through fits only to low mass stars.  

The fit to the lower envelope is denoted by the blue curve in Figure \ref{fig:clusters}. The curve is unable to match the data at the lowest masses (recall, however, that the fitting procedure for the slow launch model does not include object below $0.8 M_{\odot}$; see Section \ref{sec:methods_calibration}). A similar difficulty was noted in \citet{sills2000}: the lowest mass stars in cluster appear to be poorly fit by standard loss laws. More important for our study, however, are stars with masses $\gtrsim 1.0$. The slow and fast launch curves converge for masses $\sim 1.0$, and produce a range of predicted rotation periods for higher masses. The two curves reproduce a reasonable range of rotation rates suggested by the period data from M37 and $v\sin{i}$ data from the Hyades and Praesepe for the hot stars.

\subsubsection{Agreement with Old Clusters} \label{sec:old_clusters}
\begin{figure*}
	\centerline{\includegraphics[scale = 1.0]{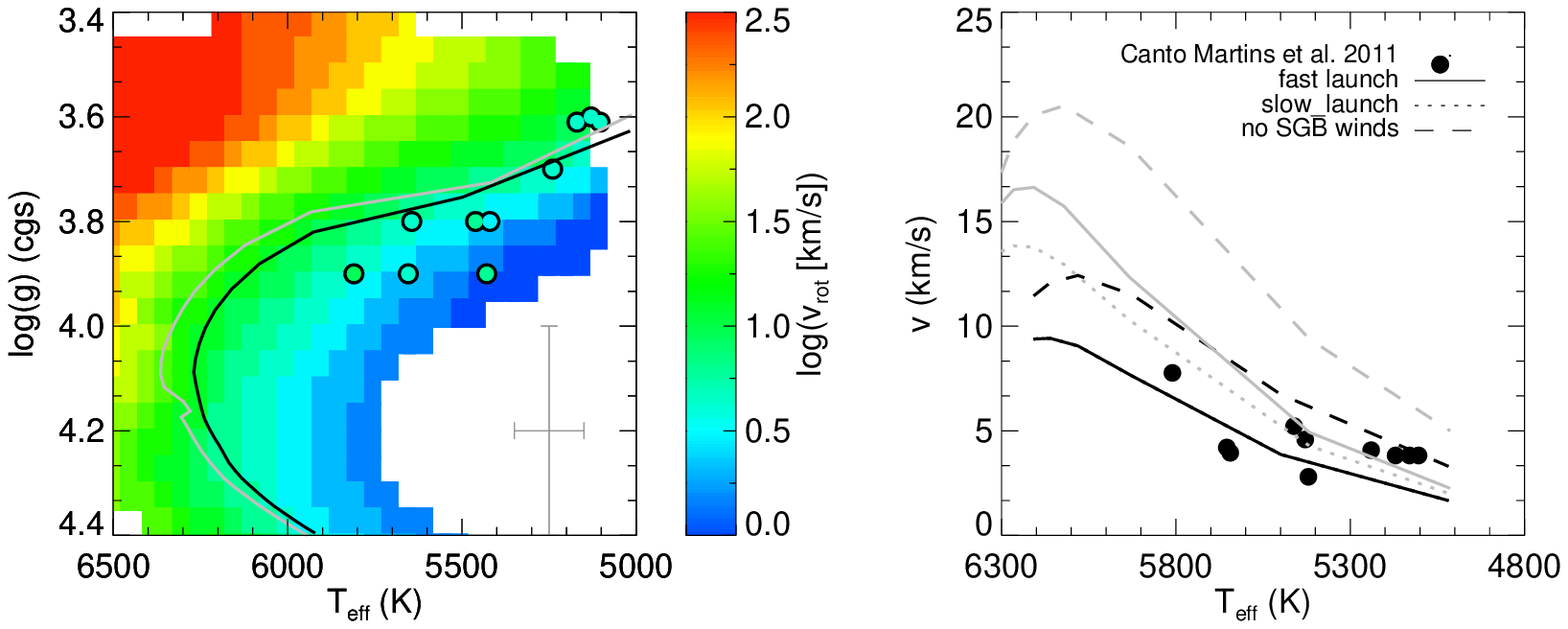}}
        \caption{Left panel: Theoretical rotation models on a $\log{g}-T_{eff}$ diagram, overplotted with the M67 data from \citet{cantomartins2011}. Color encodes rotational velocity, with the reddest colors denoting rapid rotation, for both theoretical contours and observational data. Data $v\sin{i}$ values are multiplied by a factor of $4/\pi$ to account for average inclination effects. Isochrones for a 4.0 Gyr and 3.5 Gyr solar composition population are shown in solid black and gray curves, respectively.  In the lower right is a typical errorbar for the cluster data. Right panel: Comparison of theoretical rotational isochrones (curves) and observational data (points, corrected by a factor of $4/ \pi$). Black curves denote 4.0 Gyr isochrones for fast launch (solid), slow launch (dotted), and no SGB wind (dashed). }
	\label{fig:SGB_M67}
\end{figure*}

Old open clusters have the potential to be extremely useful for validating our theoretical predictions, since the stellar parameters of cluster members can be well determined. However, there are very few old open clusters, and even fewer with good rotation data. In light of this, we compare our models to data from a single old open cluster. M67 is old enough \citep[3.5-4.0 Gyr;][]{sarajedini2009} to have a substantial number of stars near the Kraft break present on the subgiant branch, and the analysis of \citet{cantomartins2011} provides $v\sin{i}$ measurements for cluster members. In Figure \ref{fig:SGB_M67} we show our theoretical models overplotted with the cluster $v\sin{i}$'s from objects classified as subgiants or giants in \citeauthor{cantomartins2011}, and corrected by $4/ \pi$ to account for mean inclination effects. 

We perform a more quantitative test of the agreement between the data and theory for M67. We select a subset of the data that falls on the subgiant branch or early giant branch (roughly $\log{(g)} \ge 3.5$). In the CMD, the subgiant branch is narrow, indicating that the observed $\log{g}$ dispersion is observational scatter. We therefore assign isochrone positions based on $T_{eff}$ alone. 

To test whether our model predictions produce a distribution of rotation rates consistent with those found in M67, we perform the following test. We sample the M67 subgiant distribution with replacement. For each star we determine its location on a 4.0 Gyr rotational isochrone based on its measured effective temperature alone, and then draw from a distribution of inclination angles uniform in $\cos{i}$ and assign a $\sin{i}$ value to each source. The model M67 data is then compared to the real $v\sin{i}$ data using a two sample Kolmogorov-Smirnov test. We repeat this process 1000 times (sampling the distribution, assignment of $i$, and comparison to data) to obtain a reasonable average $p$ value, and well as the range of $p$ for the slow launch, fast launch, and no SGB wind models. 

In the case of M67 we also have additional information from Li studies. The work of \citet{balachandran1995} suggests that the Li is depleted in the subgiants of M67, and that the depletion patterns require that all stars on the SGB originate from ZAMS temperatures in the range of $6300-6600$ K, the so-called ``lithium dip''. Our 4.0 Gyr rotational isochrones originate from models with ZAMS effective temperatures in the $6400-6600$ K range \citep[also consistent with the modeling in][]{cantomartins2011}, and we are therefore assured that any differences between the observational and theoretical rotation distributions are due to our treatment of the rotational evolution, not because we are comparing the stars with models that originated from different places on the HR diagram.  

We find a median $p$ value of 0.11 for models using a 4.0 Gyr old solar metallicity isochrone under the fast launch conditions, suggesting that the observed and model distributions are consistent with being drawn from the same population. The slow launch case yields a similar result, with $p = 0.31$. In comparison, if we consider our models with no winds on the SGB (discussed in \ref{sec:nowind}) we find  $p = 0.001$, and that the two samples are inconsistent with having been drawn from the same parent distribution (discussed in more detail in Section \ref{sec:nowind}). If we consider instead a 3.5 Gyr isochrone, we find $p$ values of 0.03 (fast launch), 0.11 (slow launch), and $1.9 \times 10^{-5}$ (no winds). The data and rotational isochrones are shown in Figure \ref{fig:SGB_M67}. We also find, however, that the calculated $p$ values vary considerably between realizations of the sample. One must be careful not to overinterpret our results; our analysis suggests only that our rotation models are in good general agreement with the observations. There are likely to be synchronized binary and blue straggler contaminants even in this sample of subgiants and the sample size is small, which also encourages one not to place undue significance on these statistical tests.

\subsubsection{Agreement with Field Data}
\begin{figure*}
	\centerline{\includegraphics[scale = 1.0]{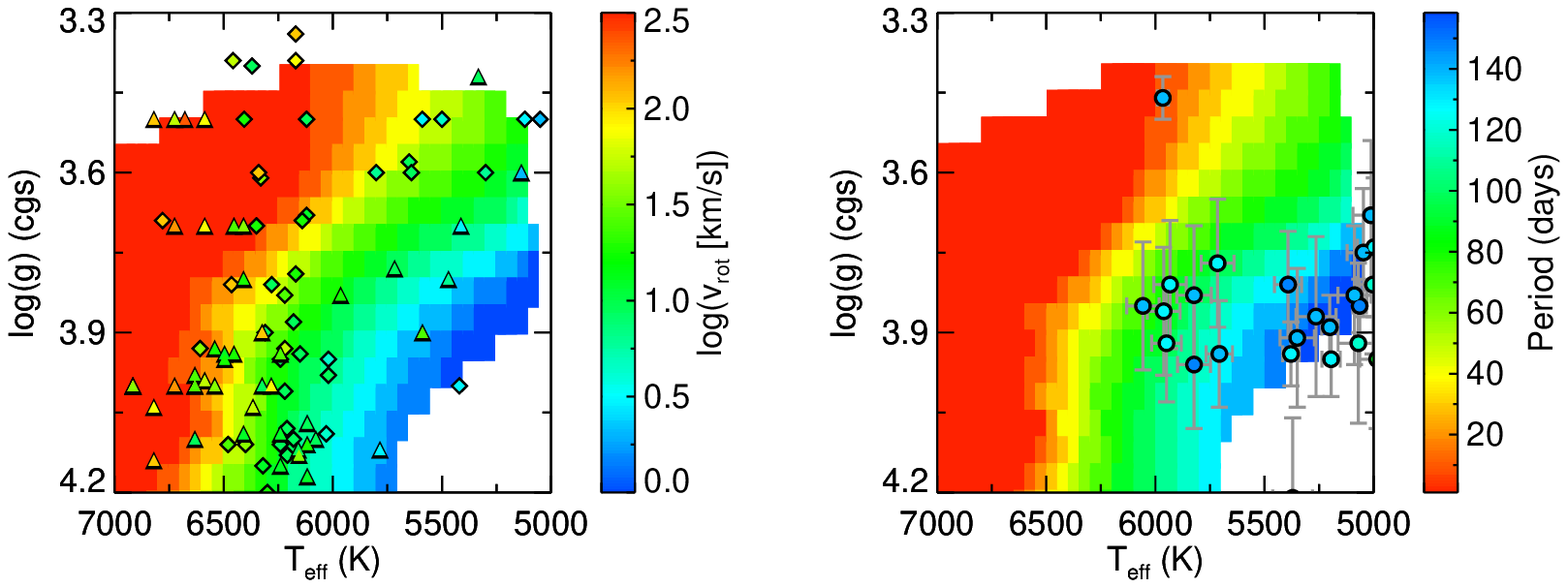}}
        \caption{Left panel: Rotation velocity is encoded by color on our theoretical log(g)-$T_{eff}$ diagram for solar metallicity. Overplotted are $v\sin{i}$ measurements for field stars from \citet{lebre1999} (diamonds) and \citet{mallik2003} (triangles), corrected by $4/\pi$ to account for average inclination effects. Right panel: Rotation period in days is encoded by color on a $\textrm{T}_{eff}$ vs. log(g) plot of our theoretical models at solar composition. Over plotted are the values of $\textrm{P}_{\textrm{rot}}$ measured by \citet{donascimento2012} with stellar parameters from \citet{gazzano2010}. }
	\label{fig:teffvlogg_tile_corot}
\end{figure*}

We can also compare our models to rotation data gathered for field stars. In the case of a field population the stellar parameters are in general more poorly known, and we frequently have no information on the compositions of the objects (which, as we show in Section \ref{sec:metallicity}, has some bearing on the predicted rotation periods). Nevertheless, rotation data, $\log{g}$, and $T_{eff}$ determinations for field stars exist, and we compare the results of selected studies to our models in Figure \ref{fig:teffvlogg_tile_corot}.

The right panel of Figure \ref{fig:teffvlogg_tile_corot} shows field $v\sin{i}$ data from \citet{lebre1999} and \citet{mallik2003}, corrected by $4/ \pi$ for the average inclination effect. The \citeauthor{lebre1999} sample is largely drawn from a collection of subgiants in the Bright Star Catalog, and the \citeauthor{mallik2003} from Hipparcos stars. As in the case of M67, the agreement between the models and the data is fair when one considers that these are lower limits on the rotation velocities. There are points with slower rotation velocities than we predict (which can be the result of $\sin{i}$ ambiguity), and very few points that are rotating substantially faster than we predict (which cannot be explained by invoking $\sin{i}$). We do not proceed with more detailed modeling due to the additional uncertainties introduced by star formation history and the range of metallicities found in a field population. 

The left-hand panel shows the results of the \citet{donascimento2012} analysis of light curves from the \textit{CoRoT} spacecraft for a second sample of field stars. Periods are determined through spot modulation, and stellar parameters are drawn from \citet{gazzano2010}. Here, unlike in M67, there are clear differences between the data and the models, namely that the highest $\textrm{T}_{eff}$ objects around 6000 K appear to be rotating more slowly than expected, and objects at 5000-5500 K too rapidly. There are (at least) two possible explanations for this disagreement. There may be deficiencies in our models, although it is unclear why such deficiencies would not be evident in the $v\sin{i}$ distributions in both the field and M67. There could also be systematic errors in the measurements of the periods and stellar parameters themselves (notably the effective temperatures). The \textit{CoRoT} light curves extend only to 150 days, while the majority of reported periods are 60 days or longer, meaning that the time series can capture at maximum two full rotation periods. The longest periods that we predict for our cool subgiants are inaccessible with time series of this length. It is unclear whether large samples of rotation periods (rather than $v\sin{i}$'s) will show similar discrepancies. In either case, this is example of the utility of a set of predictions for rotation periods as a function of mass and evolutionary state: it can either serve as a consistency check, or a test of our AM transport models.

\subsection{Rotational Tracks} \label{subsec:isochrones}
We present our rotation predictions as a function of gravity and effective temperature in Tables \ref{tbl:teff_kepz_ms}, \ref{tbl:teff_kepz_sgb}, \ref{tbl:teff_solz_ms}, \ref{tbl:teff_solz_sgb}. Table \ref{tbl:teff_kepz_ms} provides rotation periods and stellar parameters for an emsemble of MS models with masses $ 0.4 < M/M_{\odot} < 2.0 $, ages $0.55 < t < 10.0$ Gyr, and metallicity of $\textrm{[Fe/H]} = -0.2$, interpolated onto an even grid in $\log{g}$ and $T_{eff}$. Tracks for high mass models are truncated before the hook at the end of the MS to avoid the ambiguity in interpolation. For effective temperatures of 6550 K and 6600 K (6500 K) and metallicity $\textrm{[Fe/H] = -0.2}$ ($\textrm{[Fe/H] = 0.0}$) the grids are truncated with $\log{g} > 4.15$ ($\log{g} > 4.2$) to avoid regions on the HR diagram susceptible to ambiguous interpolations where the MS tracks overlap even after the hooks are excluded. Table \ref{tbl:teff_kepz_sgb} provides the same information for models on the SGB (defined as $X_c < 0.0002$). Tables \ref{tbl:teff_solz_ms} and \ref{tbl:teff_solz_sgb} are structured in the same fashion, but provide model predictions for solar metallicity.  

\begin{deluxetable*} {lllllllll}
\tabletypesize{\scriptsize}
\tablecolumns{9}
\tablewidth{0pt}
\tablecaption{Rotational tracks at fixed $T_{eff}$, [Fe/H] = -0.2, MS only}
\tablehead{
		        \colhead{$T_{eff}$ (K)}		                 &
			\colhead{log(g)}				 &
                        \colhead{Mass $\textrm{M/M}_{\odot}$}            &
                        \colhead{$\textrm{log(L/L)}_{\odot}$}            &
		        \colhead{ ZAMS $T_{eff}$ (K)}		         &
			\colhead{$P_{fast}$} 				 &
			\colhead{$P_{slow}$} 				 &
			\colhead{$I_{tot}$ $\textrm{g cm}^{-2}$}	 &
			\colhead{$\tau_{cz}$ (s)}			 }

\startdata
5000 & 4.590 & 0.7514 & -0.5270 & 4865 &  36.1167 &  36.8401 &  3.9849E+53 &  1.6491E+06 \\ 
5000 & 4.595 & 0.7562 & -0.5293 & 4888 &  32.9830 &  33.7554 &  4.0250E+53 &  1.6422E+06 \\ 
5000 & 4.600 & 0.7611 & -0.5314 & 4912 &  29.6325 &  30.4686 &  4.0665E+53 &  1.6353E+06 \\ 
5000 & 4.605 & 0.7662 & -0.5336 & 4937 &  26.0318 &  26.9548 &  4.1091E+53 &  1.6283E+06 \\ 
5000 & 4.610 & 0.7712 & -0.5357 & 4959 &  22.0552 &  23.1089 &  4.1520E+53 &  1.6212E+06 \\ 

\enddata
\label{tbl:teff_kepz_ms}
\tablecomments{This table is available in its entirety in a machine-readable form in the online journal. A portion is shown here for guidance regarding its form and content. $P_{fast} $ gives the predicted rotation period from the fast launch models, $P_{slow}$ the prediction from the slow launch case. $I_{tot}$ is the moment of inertia, and $\tau_{cz}$ the convective overturn timescale.}
\end{deluxetable*}

\begin{deluxetable*} {lllllllll}
\tabletypesize{\scriptsize}
\tablecolumns{9}
\tablewidth{0pt}
\tablecaption{Rotational tracks at fixed $T_{eff}$, [Fe/H] = -0.2, SGB only}
\tablehead{
		        \colhead{$T_{eff}$ (K)}		                 &
			\colhead{log(g)}				 &
                        \colhead{Mass $\textrm{M/M}_{\odot}$}            &
                        \colhead{$\textrm{log(L/L)}_{\odot}$}            &
		        \colhead{ ZAMS $T_{eff}$ (K)}		         &
			\colhead{$P_{fast}$} 				 &
			\colhead{$P_{slow}$} 				 &
			\colhead{$I_{tot}$ $\textrm{g cm}^{-2}$}	 &
			\colhead{$\tau_{cz}$ (s)}			 }

\startdata
5000 & 3.500 & 1.2626 & 0.7721 & 6634 &  86.8477 &  97.9382 &  1.6463E+55 &  5.2812E+06 \\ 
5000 & 3.505 & 1.2616 & 0.7732 & 6644 &  86.9717 &  98.0782 &  1.6486E+55 &  5.2888E+06 \\ 
5000 & 3.510 & 1.2605 & 0.7743 & 6653 &  87.0958 &  98.2181 &  1.6510E+55 &  5.2963E+06 \\ 
5000 & 3.515 & 1.2553 & 0.7708 & 6644 &  88.6886 &  99.4228 &  1.6267E+55 &  5.2732E+06 \\ 
5000 & 3.520 & 1.2469 & 0.7629 & 6618 &  91.4459 & 101.5148 &  1.5795E+55 &  5.2342E+06 \\ 

\enddata
\label{tbl:teff_kepz_sgb}
\tablecomments{This table is available in its entirety in a machine-readable form in the online journal. A portion is shown here for guidance regarding its form and content. $P_{fast} $ gives the predicted rotation period from the fast launch models, $P_{slow}$ the prediction from the slow launch case. $I_{tot}$ is the moment of inertia, and $\tau_{cz}$ the convective overturn timescale.}
\end{deluxetable*}

\begin{deluxetable*} {lllllllll}
\tabletypesize{\scriptsize}
\tablecolumns{9}
\tablewidth{0pt}
\tablecaption{Rotational tracks at fixed $T_{eff}$, [Fe/H] = 0.0, MS only}
\tablehead{
		        \colhead{$T_{eff}$ (K)}		                 &
			\colhead{log(g)}				 &
                        \colhead{Mass $\textrm{M/M}_{\odot}$}            &
                        \colhead{$\textrm{log(L/L)}_{\odot}$}            &
		        \colhead{ ZAMS $T_{eff}$ (K)}		         &
			\colhead{$P_{fast}$} 				 &
			\colhead{$P_{slow}$} 				 &
			\colhead{$I_{tot}$ $\textrm{g cm}^{-2}$}	 &
			\colhead{$\tau_{cz}$ (s)}			 }

\startdata
5000 & 4.570 & 0.7993 & -0.4802 & 4886 &  39.9515 &  40.7282 &  4.6844E+53 &  1.7751E+06 \\
5000 & 4.575 & 0.8042 & -0.4825 & 4907 &  36.8166 &  37.6408 &  4.7270E+53 &  1.7679E+06 \\ 
5000 & 4.580 & 0.8092 & -0.4848 & 4930 &  33.4732 &  34.3580 &  4.7715E+53 &  1.7609E+06 \\ 
5000 & 4.585 & 0.8143 & -0.4871 & 4952 &  29.9248 &  30.8859 &  4.8174E+53 &  1.7542E+06 \\ 
5000 & 4.590 & 0.8195 & -0.4894 & 4975 &  26.0308 &  27.0990 &  4.8649E+53 &  1.7473E+06 \\ 

\enddata
\label{tbl:teff_solz_ms}
\tablecomments{This table is available in its entirety in a machine-readable form in the online journal. A portion is shown here for guidance regarding its form and content. $P_{fast} $ gives the predicted rotation period from the fast launch models, $P_{slow}$ the prediction from the slow launch case. $I_{tot}$ is the moment of inertia, and $\tau_{cz}$ the convective overturn timescale.}
\end{deluxetable*}

\begin{deluxetable*} {lllllllll}
\tabletypesize{\scriptsize}
\tablecolumns{9}
\tablewidth{0pt}
\tablecaption{Rotational tracks at fixed $T_{eff}$, [Fe/H] = 0.0, SGB only}
\tablehead{
		        \colhead{$T_{eff}$ (K)}		                 &
			\colhead{log(g)}				 &
                        \colhead{Mass $\textrm{M/M}_{\odot}$}            &
                        \colhead{$\textrm{log(L/L)}_{\odot}$}            &
		        \colhead{ ZAMS $T_{eff}$ (K)}		         &
			\colhead{$P_{fast}$} 				 &
			\colhead{$P_{slow}$} 				 &
			\colhead{$I_{tot}$ $\textrm{g cm}^{-2}$}	 &
			\colhead{$\tau_{cz}$ (s)}			 }

\startdata
5000 & 3.500 & 1.5109 & 0.8577 & 7236 &  55.5760 &  74.1348 &  2.5881E+55 &  6.1224E+06 \\ 
5000 & 3.505 & 1.5102 & 0.8589 & 7247 &  55.6554 &  74.2408 &  2.5918E+55 &  6.1312E+06 \\ 
5000 & 3.510 & 1.5039 & 0.8543 & 7224 &  56.0574 &  74.3405 &  2.5479E+55 &  6.0940E+06 \\ 
5000 & 3.515 & 1.4944 & 0.8466 & 7183 &  56.6592 &  74.2962 &  2.4789E+55 &  6.0576E+06 \\ 
5000 & 3.520 & 1.4853 & 0.8389 & 7142 &  57.3912 &  74.3488 &  2.4137E+55 &  6.0217E+06 \\ 

\enddata
\tablecomments{This table is available in its entirety in a machine-readable form in the online journal. A portion is shown here for guidance regarding its form and content. $P_{fast} $ gives the predicted rotation period from the fast launch models, $P_{slow}$ the prediction from the slow launch case. $I_{tot}$ is the moment of inertia, and $\tau_{cz}$ the convective overturn timescale.}
\label{tbl:teff_solz_sgb}
\end{deluxetable*}

\section{Stellar Rotation as a Tool} \label{sec:tool}
Because rotation is a strong function of mass and evolutionary state, and because sharp features such as a the Kraft break are present and persist in the rotation distribution, it can be used as a powerful diagnostic of the underlying stellar populations. In this section we expand traditional gyrochronology relationships to include subgiants and discuss diagnostic tests of mass and radius. We also quantify the impact of the SGB on field populations.

\subsection{Stellar Ages}
\subsubsection{Existing rotational age diagnostics}
We are accustomed to thinking of relationships between period and stellar ages in the context of gyrochronology of cool dwarfs \citep{barnes2007,mamajek2008, meibom2009, meibom2011}. While these period-age relationships have the potential to be powerful indicators of age in the field, they are applicable only for cool, unevolved stars. The gyrochronology relations rely on AM loss over time to produce a smooth transition from rapidly rotating young stars to slowly rotating old stars. The strong dependence of the loss law on the rotation rate quickly erases the memory of the initial conditions, thus producing a sequence in which all stars of a given mass at a given age rotate at the same rate. Hot stars born above the Kraft break are therefore unsuitable for gyrochronology, since they do not undergo strong wind losses, and have little relationship between their periods and ages on the MS. Likewise, evolved stars that undergo expansion and increase in their moments of inertia are also not suitable for traditional gyrochronology, which assumes that the stellar spin down is due only to AM losses. In clusters it is possible to focus only on stars suitable for gyrochronology; in a field population, these hot and evolved stars will mingle with the cool dwarfs, the effects of which we discuss below.

\subsubsection{A more complicated picture: rotational regimes for field populations}

\begin{figure*}
	\centerline{\includegraphics[scale = 1.0]{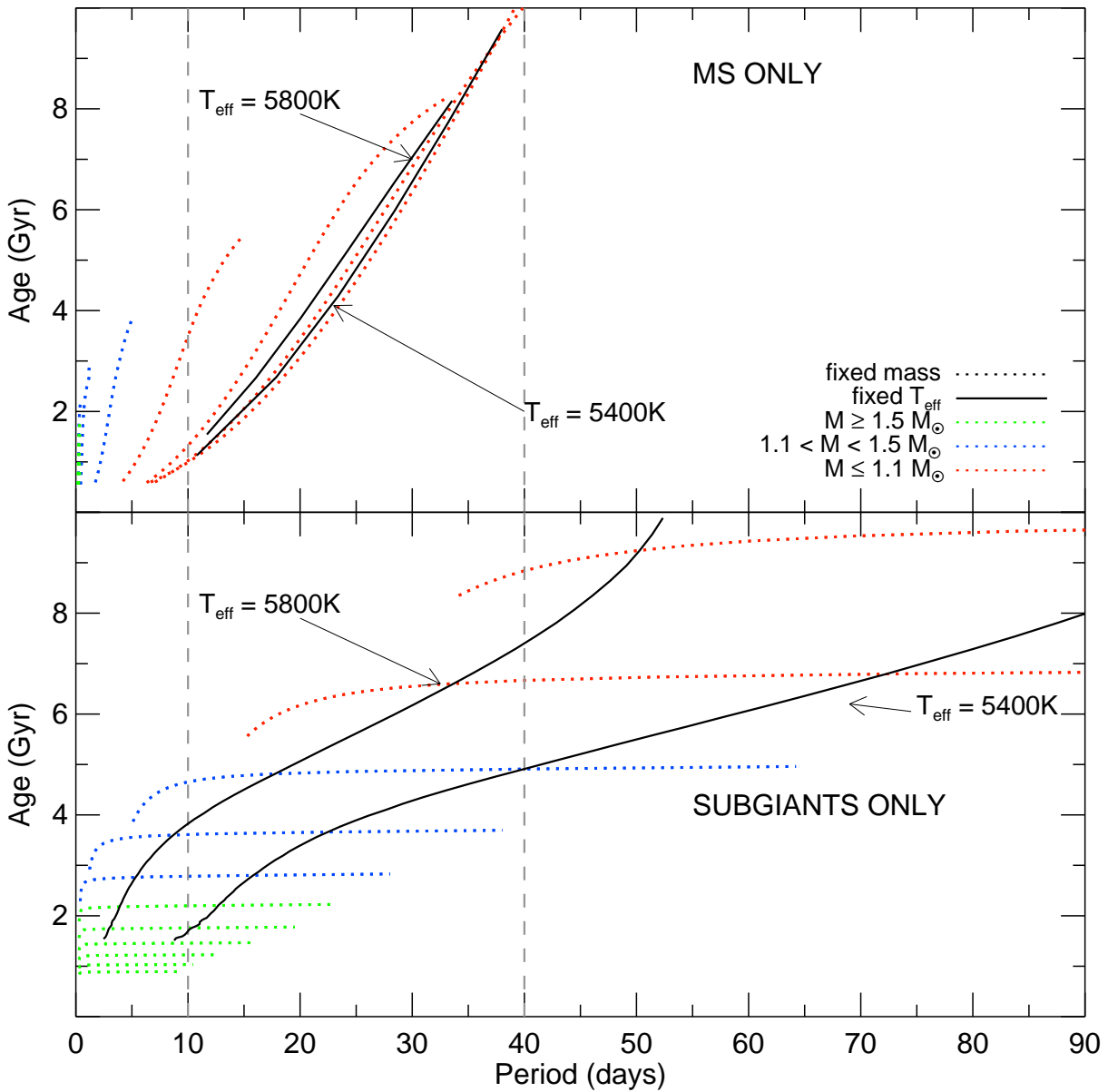}}
        \caption{Age versus period for the upper envelope in rotation periods on the MS (top panel) and SGB (lower panel). Green curves represent objects with masses $M \geq 1.5 M_{\odot}$, blue curves those with $1.1 < M/M_{\odot} < 1.5$, and red those curves with masses $M \leq 1.1 M_{\odot}$. Solid black curves are lines of constant effective temperature in the period-age plane.}
	\label{fig:periodage_MSSGB}
\end{figure*}

Figure \ref{fig:periodage_MSSGB} shows the period as a function of age for both MS stars (top panel) and subgiants (bottom panel). Also shown are slices at fixed effective temperatures (solid black curves). Cool dwarfs ($M < 1.1 M_{\odot}$) in the top panel display the tight period-age relationship that makes gyrochronology possible. Given a period, with some small correction for mass, one can infer the age. The relationship is not as clear for more massive stars ($M > 1.1 M_{\odot}$): here the curves become nearly vertical, indicating that there is little change in the period as a function of age, and that a measurement of the period does not imply a unique stellar age. This is a restatement of the fact that the gyrochronology relations are not applicable to hot stars without strong MS AM loss. 

We see in the lower panel that although subgiants may not obey the same period-age relationships as their MS counterparts, these two quantities are in fact related. If we add information about the effective temperature, in particular, we obtain tight sequences that yield unique ages at a given period for subgiants. We can understand this trend as follows: objects leave the MS with a strongly mass and age-dependent rotation rate. Massive stars are born rapidly rotating and do not spin down on the MS, and cooler stars are spun down substantially and have increasingly longer periods the older they are. SGB lifetimes are also strongly mass dependent in the same sense: more massive objects are younger upon arrival on the SGB, and spend less time there. We therefore see that more rapidly rotating stars on the SGB have younger absolute ages, whereas the most slowly rotating stars are also the oldest. This subgiant period-age relationship has not been previously appreciated in the literature. It implies that the diagnostic power of the gyrochronology relationships can be extended to more evolved stars, provided we account for the different mechanisms responsible for their period distributions and spin down. 

\begin{figure*}
	\centerline{\includegraphics[scale = 0.9]{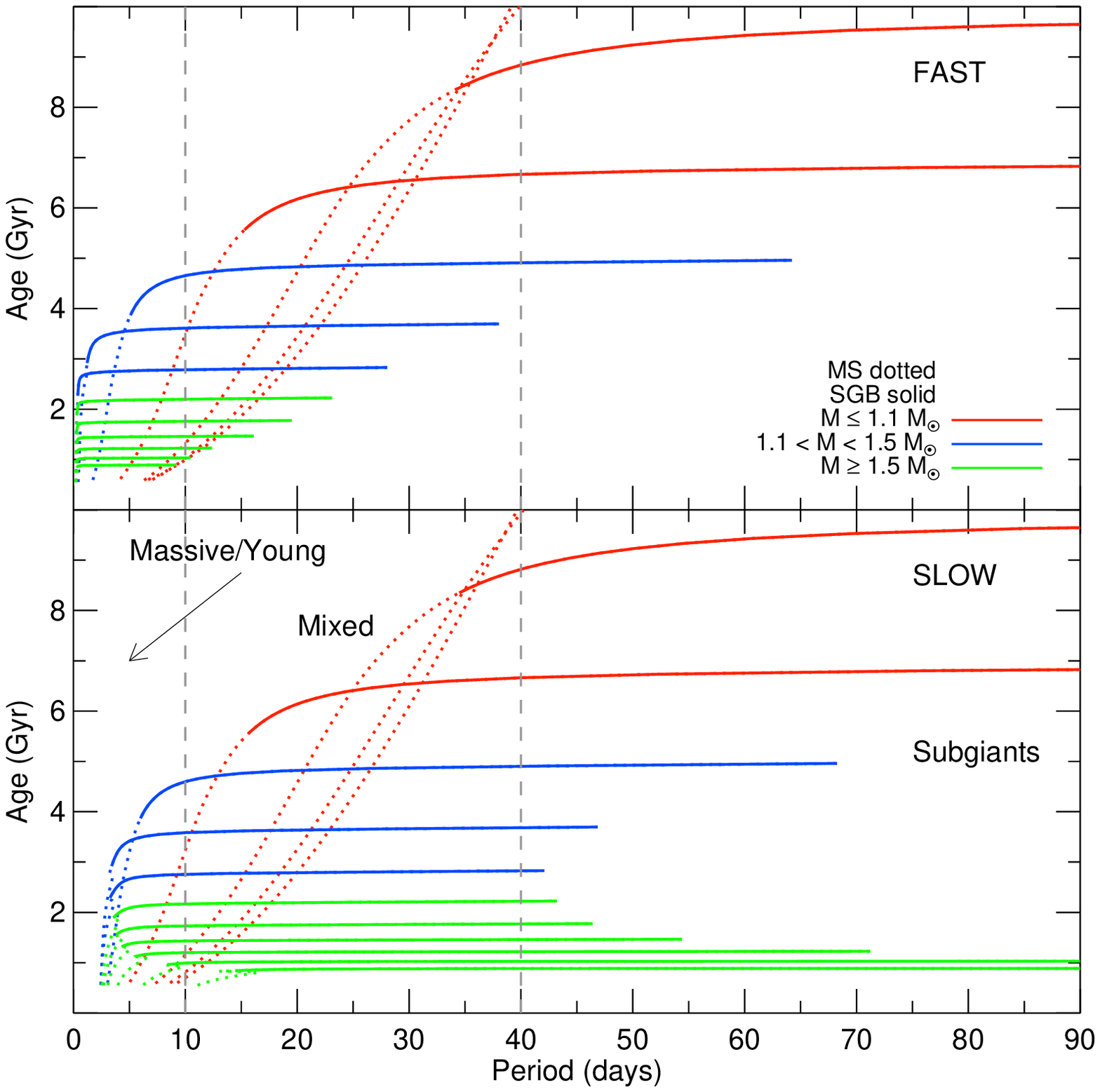}}
        \caption{Age versus period for the upper envelope in rotation periods (top panel) and initial periods an order of magnitude slower (bottom panel). Green curves represent objects with masses $M \geq 1.5 M_{\odot}$, blue curves those with $1.1 < M/M_{\odot} < 1.5$, and red those curves with masses $M \leq 1.1 M_{\odot}$. Solid colored curves are objects on the SGB, dotted curves objects on the MS.}
	\label{fig:periodage_fastslow}
\end{figure*}

In Figure \ref{fig:periodage_fastslow} we combine both the evolved and unevolved stars onto a single diagram for both the fast launch (top panel) and slow launch (bottom panel) conditions, as they would be in a field population. We divide this diagram into three distinct rotational regimes: 

\begin{enumerate}
\item{$P < 10$ days: Massive ($M > 1.1 M_{\odot}$) or young stars. All stars in this period range are either above a solar mass (on the MS or the SGB) or young, rapidly rotating solar/subsolar mass objects.}
\item{$10<P<40$ days: Mixed. Objects in this period range can be low mass MS stars, or subgiants at a range of masses.}
\item{$P > 40$ days: Subgiants. The vast majority of objects in this period range are low or intermediate mass subgiants.}
\end{enumerate}

These three regimes have corresponding designations for traditional gyrochronology: $P < 10$ days denotes young objects, $10 < P < 40$ days objects suitable for gyrochronology, and objects with $P > 40$ days are outside the valid period range for the gyro relations. When we consider a field population we must modify these traditional regimes: rapidly rotating objects can be young \textit{or} massive. Objects in the $10 < P < 40$ day are good targets for gyrochronology, but we must be very careful about assigning the correct evolutionary state to targets within this range, lest we apply the wrong period-age relationship. Objects with periods longer than 40 days are evolved, and present their own period-age relationships.

\subsubsection{The Importance of Subgiants} \label{subsubsec:importance_of_SGB}

Not only does the period distribution of subgiants overlap with that of the MS in the critical $10 < P < 40$ day range, but subgiants also represent significant contaminants in surveys of field populations. Subgiant stars will be systematically over-represented in magnitude-limited surveys. We use a TRILEGAL galaxy model with default parameter values \citep[see][and the web submission form \footnotemark ]{girardi2005,girardi2012} for a 5 $\textrm{deg}^2$ field of view (equivalent to the area covered by a single \textit{Kepler} CCD module) centered on the \textit{Kepler} field at galactic coordinates $(l,b) = (76.32^{\circ}, +13.5^{\circ})$ to estimate the fraction of subgiants. If we consider objects in a $100 K$ slice centered on $5800 K$ and define all objects in that slice with $\log(g) \geq 4.2$ to be subgiants, we find that for a limiting \textit{Kepler} magnitude of $K_p < 14$, $~35\%$ of the stars in the sample are subgiants, $73\%$ of which have masses $M > 1.1 M_{\odot}$. We obtain similar results with a limiting SDSS $g $ magnitude of $g < 14$. We show the results for different magnitude cuts in Table \ref{tbl:mag_limits}. We have not included magnitude errors in the analysis, nor tuned the TRILEGAL model to specifically match the \textit{Kepler} field, but we expect that a more careful approach will yield a qualitatively similar result: subgiants are an important contaminant in magnitude limited surveys in this temperature range, which requires that any survey seeking to use gyrochronology to determine the ages of stars in such a sample must be able to discriminate between MS and post-MS objects. The case of the two subgiants in \citet{dogan2013} is an excellent example of this phenomenon. The gyrochronology and asteroseismic ages for these objects do not match, and without the excellent gravity and age information provided by seismology, they would likely have been misclassified as old dwarfs.  Failure to identify these frequent subgiant contaminants and treat them accordingly will result in erroneous ages for more than a third of the stars in these surveys. 

\footnotetext{http://stev.oapd.inaf.it/cgi-bin/trilegal}

\begin{deluxetable} {lllll}
\tabletypesize{\scriptsize}
\tablecolumns{5}
\tablewidth{0pt}
\tablecaption{Subgiant fractions}
\tablehead{
                        \colhead{Mag. limit}                       &
                        \colhead{$\displaystyle\frac{N_{sgb}}{N_{tot}}$}                     &
			\colhead{$\displaystyle\frac{N_{P < 10}}{N_{sgb}}$}                     &
		        \colhead{$\displaystyle\frac{N_{sgb, M > 1.1}}{N_{sgb}}$}	&
			\colhead{$\displaystyle\frac{N_{sgb, M > 1.5}}{N_{sgb}}$}}
\startdata
$K_p < 14$ & 0.35 & 0.17 & 0.73 & 0.10\\
$K_p < 11$ & 0.75 & 0.67 & 1.00 & 0.67\\

\enddata
\tablecomments{For the temperature range $5750 < T_{eff} < 5850$ K, with subgiants defined as the subset of objects with $log(g) < 4.2$. $N_{sgb}/N_{tot}$ is the fraction of subgiants, $N_{P < 10}/N_{sgb}$ the fraction of subgiants with rotation periods less than 10 days, $N_{sgb, M > 1.1}/N_{sgb}$ the fraction of those subgiants with $M>1.1 M_{\odot}$, and $N_{sgb, M > 1.3}/N_{sgb}$ the fraction of subgiants with $M>1.3 M_{\odot}$}
\label{tbl:mag_limits}
\end{deluxetable}

\subsection{Period as a Constraint on Stellar Radii}
   
One of the primary science goals of the \textit{Kepler} mission is to detect transiting extrasolar planets, and statistically characterize the planet population. Because the transit method is sensitive to the ratio of the planetary and stellar radii, one must have precise stellar parameters before accurate measurements of the planet properties are possible. The stellar parameters in the \textit{Kepler} Input Catalog \citep[KIC][]{brown2011}, while adequate for their intended purpose of distinguishing giants from dwarfs, are insufficient to determine the stellar radius with the necessary accuracy. Follow-up observations are generally essential for the careful characterization of the star.

\begin{figure*}
	\centerline{\includegraphics[scale = 1.0]{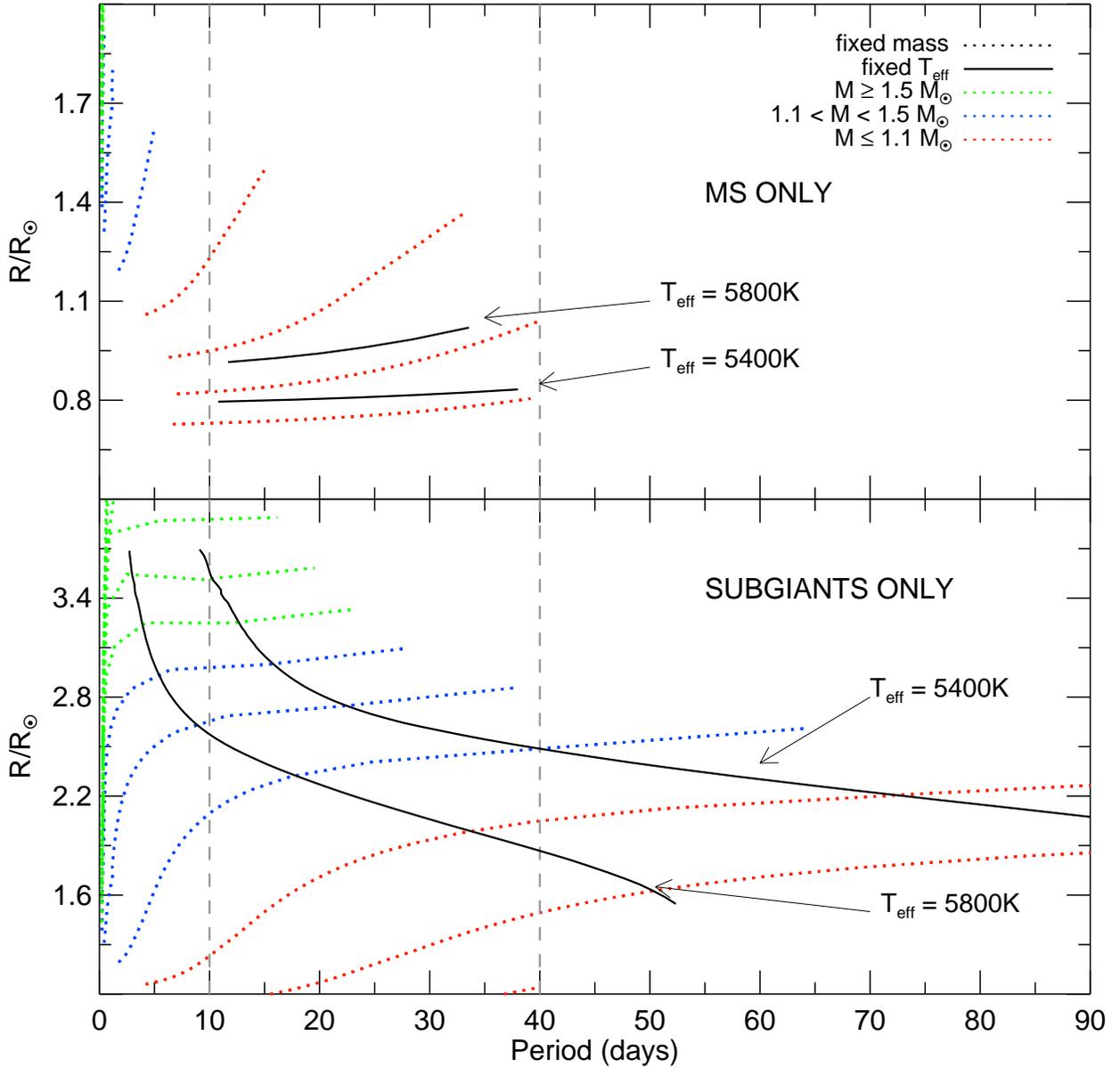}}
        \caption{Radius versus period for the upper envelope in rotation periods on the MS (top panel) and SGB (lower panel). Green curves represent objects with masses $M \geq 1.5 M_{\odot}$, blue curves those with $1.1 < M/M_{\odot} < 1.5$, and red those curves with masses $M \leq 1.1 M_{\odot}$. Solid black curves are lines of constant effective temperature in the period-radius plane.}
	\label{fig:periodradius_MSSGB}
\end{figure*}

We suggest that the stellar rotation period can also offer constraints, at least in a statistical sense, on the radii of the stars at fixed effective temperature. In Figure \ref{fig:periodradius_MSSGB} we show theoretical models for the relationship between period and radius for the fast launch conditions. Radius is not a strong function of period on the MS, because the radii of MS stars change very little over the course of their core-hydrogen burning evolution, but they undergo substantial evolution in their periods due to magnetic braking. On the subgiant branch, however, because both the increase in period and increase in radius are linked to the expansion of the stellar envelope, there exists a correlation at fixed effective temperature between the radius and the rotation period. Therefore, if one has a way of reliably distinguishing between dwarfs and subgiants (via coarse $\log{g}$ measurements or luminosity, for example), a measurement of the period with provide an additional constraint on the stellar radius.

\subsection{Period as a Test of the Mass Scale}

Because period is a strong function of stellar mass, both on the MS and the SGB, measurements of rotation periods could be instrumental in helping to distinguish between massive and low-mass objects, particularly on the subgiant branch. Spectroscopic gravity measurements have notoriously large uncertainties, often of the order of 0.5 dex, which severely hampers classification of stars on the subgiant branch. The addition of a rotation measurement can help to distinguish between masses on the subgiant branch, provided that the determination of $\log{g}$ is sufficient to rule out a MS dwarf. 

Using the same TRILEGAL model discussed in Section \ref{subsubsec:importance_of_SGB}, we show that one could systematically select massive subgiants for study based on their rotation periods. If we use our ``fast launch'' model grid to define the regions in a $\log{g}-T_{eff}$ diagram populated by rapid rotators, and again consider a narrow slice in temperature of 100 K around 5800 K, we find that $24\%$ of subgiants should have $P < 10$ days. Figure \ref{fig:periodage_MSSGB} shows that for the fast launch, all subgiants in this period range are also relatively massive. If we consider the entire long-cadence \textit{Kepler} sample with effective temperatures from \citet{pinsonneault2012}, $\sim9300$ stars fall within this effective temperature slice, which would imply a sample of $\sim 800$ massive, rapidly rotating subgiants available for study.

\section{Discussion} \label{sec:discussion}

We have necessarily made simplifying approximations in our models. We have considered rotation across a population, but for that population we have assumed solid body rotation, a single composition, limited stellar demographics, and simplified stellar models. In this section we discuss and quantify the impact of such assumptions. We quantify the importance of both metallicity and uncertain processes such as winds on the SGB for the model predictions of rotation periods in Section \ref{subsec:parameter_variations}. In Section \ref{sec:complications} we discuss the implications of a more realistic stellar population that includes binaries and other subgroups with unusual rotation distributions, as well as the impact of some of our simplified modeling approaches on our results.  

Finally, in Section \ref{sec:add_tools} we also speculate about the use of rotation as a tool in the future. We discuss tests of the physics of angular momentum transport via measurements of the surface rotation rates, as well as using rotation to verify the asteroseismic mass scale. We close in Section \ref{sec:gaia} with a discussion of the diagnostic power of rotation in the era of precision parallaxes from missions such as Gaia.

\subsection{Parameter Variations} \label{subsec:parameter_variations}

We present in this section the results of our parameter variation studies, which provide insight into the importance of physical processes such as continuing AM loss on the SGB through winds, as well as addressing the importance of metallicity to our conclusions. 

\subsubsection{Winds on the SGB} \label{sec:nowind}

Hot stars experience minimal AM loss on the MS. However, our diagnostics of stellar activity indicate that subgiants possess active chromospheres and coronae. Studies such as \citet{schrijver1993} also argue for the presence of winds based on rotation periods. We therefore expect that massive subgiants will spin down once cool, and have incorporated this into our models. However, the behavior and strength of subgiant AM loss remains uncertain. To quantify the impact of winds SGB on the periods predicted from our models, we compare models run with and without SGB AM loss. 

\begin{figure*}	
	\centerline{\includegraphics[scale = 1.0]{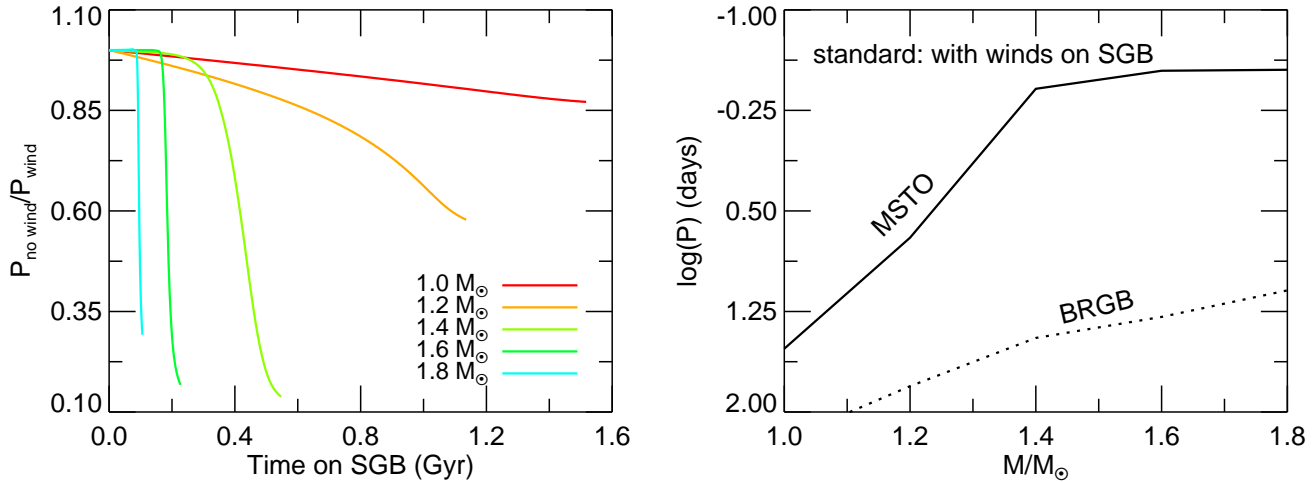}}
        \caption{Left panel: $P_{ \textnormal{ \tiny no wind}}/P_{ \textnormal{\tiny wind}}$ is the ratio of the periods for models without SGB winds to the standard case, which includes SGB wind losses. Mass is given by the color of the curves, with redder color representing lower mass models. Right panel: Period as a function of mass at both the MSTO and BRGB for the standard, fast launch models with SGB wind losses.}
	\label{fig:wind_compare_model_by_model}
\end{figure*}

In Figure \ref{fig:wind_compare_model_by_model} we plot the ratio of the periods of the no-wind models and those from the standard models as a function of time on the SGB. We use models evolved with the fast launch initial conditions where the effects of winds will be most significant. For stars born below the Kraft break (the red curve, for example) objects arrive on the SGB slowly rotating, and have convective envelopes for their entire MS and SGB lifetimes. SGB wind losses are steady, as a result of this slow rotation. The timescale for AM loss $t_{loss} \propto J_{MSTO}/<dJ/dt>$ for the models with winds is much longer than the SGB lifetime, $t_{SGB}$ ($\sim 12$ Gyr versus $\sim 1.5$ Gyr for a $1.0 M_{\odot}$ model). Therefore, despite the fact that the SGB phase is relatively long and objects undergo AM loss over the entire SGB, the difference between models with and without winds is modest. 

Objects born above the Kraft break only develop substantial CZs after arrival on the SGB, which is the reason for the sudden decline in rotation rate for the most massive stars in Figure \ref{fig:wind_compare_model_by_model}. Models with and without wind losses on the SGB are identical until convective envelopes develop and winds begin to operate. Because these stars were rapidly rotating on the MS, the appearance of a surface convection zone leads to a strong spin down. However, these stars also have the shortest SGB lifetimes of the objects considered here (0.1 Gyr for a $1.8 M_{\odot}$ model), limiting the total AM that can be lost.

Stars in the vicinity of the Kraft break (green curves in Fig. \ref{fig:wind_compare_model_by_model}) are the most strongly affected by SGB winds by the time they arrive at the base of the giant branch, due to the combination of relatively rapid rotation and long SGB lifetimes. Here $t_{loss} \sim t_{SGB}$, allowing for significant losses. These stars therefore undergo strong braking as their convective envelopes deepen, and have significant SGB crossing times over which to lose AM. These objects are the most strongly affected by the assumptions regarding winds on the SGB. 

The effects of wind losses on the SGB are substantial enough that observational datasets should be able to motivate whether or not their inclusion is appropriate. Our preliminary analysis in Section \ref{sec:old_clusters} for the stars in M67 suggests that the inclusion of winds is necessary, although there are additional uncertainties in the stellar physics that could mimic the long periods of a population of subgiants affected by winds. 

\subsubsection{Metallicity} \label{sec:metallicity}

We have considered mono-abundance model populations, but in reality, objects in a field population will support a range of compositions. Traditional stellar diagnostics have well-known degeneracies between composition, age, and mass, and we therefore investigate the sensitivity of rotation-based stellar diagnostics to composition.

Changes in metallicity affect models in two ways that are important for rotation: lower metallicity models have both shorter total lifetimes, and shallower convective envelopes. Both of these effects tend to decrease the rotation period. Winds are less effective and therefore drain less AM, and the models have shorter lifetimes over which to lose AM. We therefore expect that at fixed mass and age, the differences between models of different metallicity can be large. 

\begin{figure*}
	\centerline{\includegraphics[scale = 0.95]{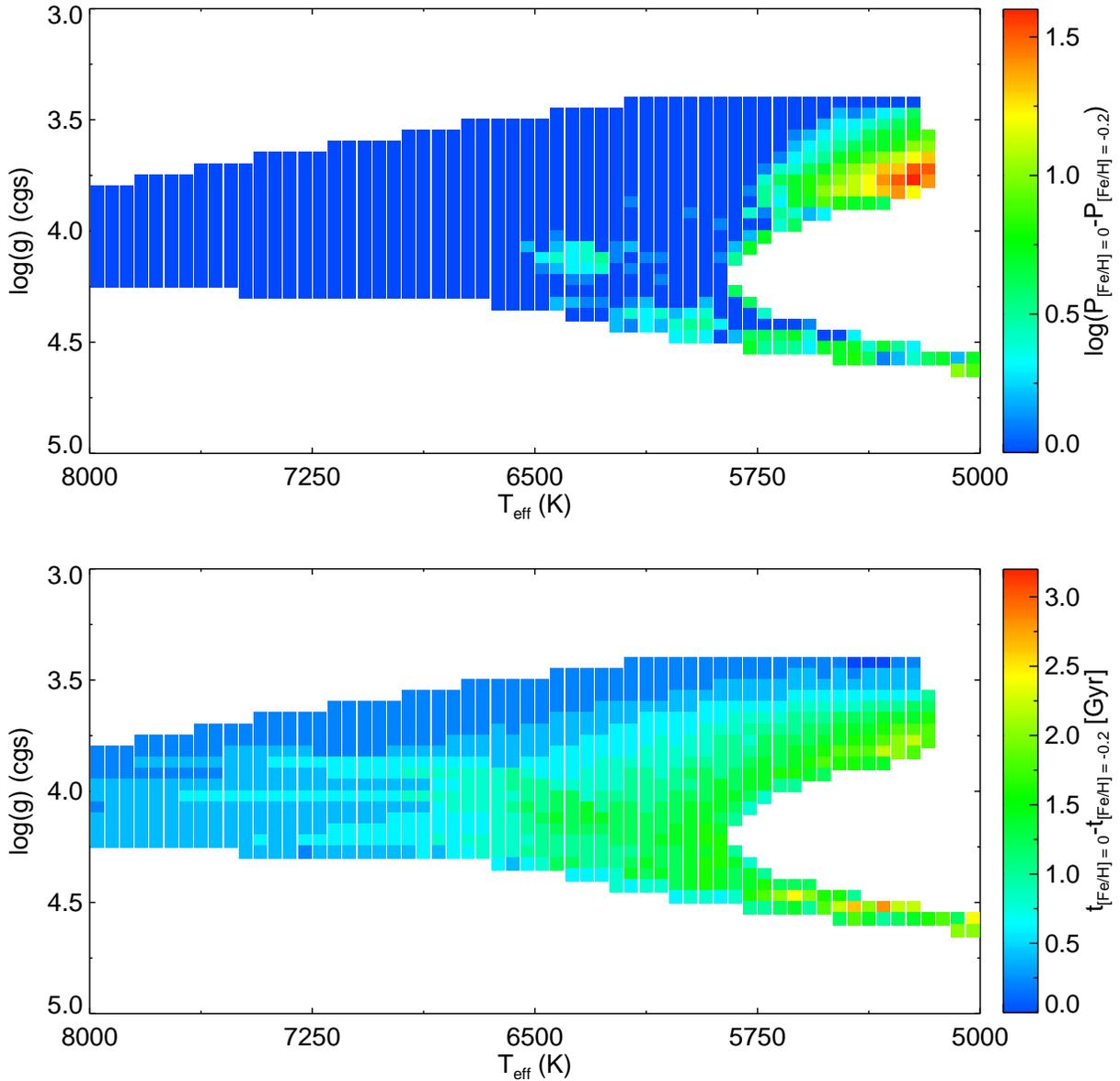}}
        \caption{Top panel: The logged absolute value of the difference in period between models at solar metallicity and $\textrm{[Fe/H] = -0.2}$ at fixed combinations of $\textrm{T}_{eff}$ and log(g). All models were generated with identical starting conditions and AM evolution schemes. Bottom panel: the difference in age between metal-rich and metal-poor models at fixed log(g) and $\textrm{T}_{eff}$.}
	\label{fig:met_compare}
\end{figure*}

However, if we view the impact of metallicity in terms of observables, namely $\log{g}$ and $T_{eff}$, the differences in the periods induced by composition are somewhat muted. Because rotational behavior is closely linked to effective temperature, we expect that the differences between the two compositions viewed in this plane will be smaller, although not zero, due to age differences. Figure \ref{fig:met_compare} shows the difference in rotation period as a function of surface gravity and effective temperature between models at solar composition and those at $\textrm{[Fe/H]} = -0.2$. We construct this diagram in a similar fashion to Figure \ref{fig:teffvlogg_colors}, except that each tile is now the difference between the average model period in a given $\log{g}-T_{eff}$ box for a solar and $\textrm{[Fe/H]}=-0.2$ grid. As expected, models with higher metallicities tend to rotate more slowly than those of lower metallicities. On average, the low metallicity models at any given location on the diagram are younger, as shown in the bottom panel of the same figure. 

The period differences are most severe for objects at the base of the giant branch; at a fixed effective temperature, a more metal rich object has expanded more than a metal poor one due to metallicity dependent shifts in the location of the Hayashi track. The lower MS, while unevolved, contains old objects that have had a long time to accumulate differences in their periods due to age and CZ depth differences. Finally, there is a discrepant patch at $\sim 6300$ K due to the fact that the appearance of hooks in the stellar tracks (due to convective cores) is shifted to lower temperatures at higher metallicities, and the morphology of these hooks affects the exact pattern of AM loss, since objects effectively cross from one side of the Kraft break to the other and back over the course of their evolution through the hook. In this region one is comparing stars that do and do not undergo evolution through a hook, and therefore have different rotation periods.

\subsection{Other Complications} \label{sec:complications}

\subsubsection{Astrophysical Backgrounds}
We have considered reasonable distributions of rotation periods for single stars on the MS, and quantified a reasonable range in period. We have, not, however, accounted for several classes of astrophysical background sources that will inevitably be present in large datasets of rotation periods:

\begin{enumerate}
 \item{Synchronized binaries: Tidally synchronized binary systems will have rotation periods less than 10 days, and they are relatively common.  \citet{duquennoy1991}, for example, found binaries with periods less than 10 days to be $4\%$ of the sample in the field, compared with 11\% in the Hyades. These objects will contaminate the rapidly rotating sample. RV variability or SED information from photometry should permit us to quantify and distinguish this population from massive single stars. }

\item{Stellar mergers: The rotation rates of mergers are not radically distinct from those of other massive stars \citep[see for example][for the old open cluster NGC 188]{mathieu2009}. For low mass stars, however, mergers could produce unusually rapid rotation for stars of the same mass and age. \citet{andronov2006} estimated that $\sim3\%$ of sub-turnoff stars in old open clusters could be merger products, and such stars would rotate more rapidly on average than single stars.  A comparable fraction of halo dwarfs are severely over-depleted in Li \citep{thorburn1994}, which is consistent with this estimate. Li abundances, for some mass ranges, should allow us to discriminate between normal and blue straggler stars. } 

\item{Slowly rotating massive stars: There are two mechanisms that can produce slow rotation in massive stars \citep[see][for a good discussion]{abt1995}. A sub-population of stars have very high magnetic fields \citep{babcock1958}, which leads to very slow surface rotation.  These stars have peculiar spectra caused by element segregation in their outer layers \citep{michaud1970}.  Tidally synchronized binaries will also rotate unusually slowly.  Such stars also are subject to unusual surface abundance patterns and are typically referred to as Am stars \citep{titus1940}. Both phenomena appear at surface rotation rates below 90-120 km/s.  Stars with an angular momentum history different from that in our models will therefore be reasonably common in massive stars. \citeauthor{abt1995} find that $16\%$  of the overall population of A0-A1 stars fall under one category or another.  The frequency of chemically peculiar stars rises into the early F domain.  The most likely explanation is not a mass dependent trend in frequency, but instead the onset of slower surface rotation, which in turn induces unusual surface abundances.}

\end{enumerate}

Each of these classes of objects represents both a background and an opportunity: while they will contaminate any large sample of rotation periods and make interpretation of those periods more challenging, we may also be able to use their unusual rotation rates to identify them as interesting subsets of objects for further study. 

\subsubsection{Modeling Considerations}
We have neglected several physical effects in our models that could be important if we intend to utilize stellar rotation as a precise diagnostic of stellar parameters. Apart from the uncertainties in the modeling of the AM loss and evolution that we have already discussed, we have also neglected the effects of stellar rotation on the structure itself, and considered models with no diffusion, both of which may be important for stars at and above the Kraft break. 

Rotation reduces the effective gravity of the star, and rotating stars have lower effective temperatures and luminosities than non-rotating stars of the same mass and composition. For a $1.6 \textrm{ M}_{\odot}$ model run in YREC's fully rotating configuration, with solid body rotation and no wind losses and a ZAMS rotation velocity of $\sim150$ km/s, the corrections to the effective temperature and gravity at the MS turnoff are within $\Delta T_{eff} = 100-300$ K throughout the MS and SGB, with $\Delta \log{g} < 0.03$ on the MS, $\Delta \log{g} < 0.07$ on the SGB. Rotation has substantially weaker effects on the structures of low mass stars \citep[see][]{sills2000}. These effects will need to be taken into account before the mapping between temperature and mass is trustworthy, but the general trends that we focus on throughout this paper regarding mass, evolutionary state, and period should remain qualitatively unchanged. 

Diffusion and gravitational settling are another class of physical effects that we have neglected in our treatment that may induce uncertainty into relationships between $T_{eff}$, $\log{g}$, age, and rotation period. Diffusion has the effect of deepening the convection zone and shortening the lifetime of a model with respect to the no diffusion case, both of which, as we have shown for metallicity, affect the predicted rotation period. 

\subsection{Additional Tools} \label{sec:add_tools}

In Section \ref{sec:tool} we discussed the many ways in which rotation can be used as a tool to understand stellar populations. In this section we discuss additional applications of rotation rates that may represent interesting tools in the near future. 

\subsubsection{Rotation as a Test of the Asterosesimic Mass Scale}

Measurements of rotation period could serve as useful diagnostics in the context of asteroseismology. There are two primary diagnostics of the global oscillation pattern in asteroseismology can provide information on the mass and radius of the star. The frequency of maximum power, $\nu_{max}$, is the location of the peak (in power) of the envelope of oscillation modes. The large frequency separation, $\Delta \nu$, is  the spacing in frequency between modes of same degree and consecutive radial order. Together, measurements of these quantities in conjunction with the scaling relations \citep{kjeldsen1995}:
\begin{equation}
 \displaystyle \frac{\Delta \nu}{\Delta \nu_{\odot}} = \left(\frac{M}{M_{\odot}}\right)^{1/2} \left(\frac{R}{R_{\odot}}\right)^{-3/2}
\end{equation}

\begin{equation}
 \displaystyle \nu_{max} = \frac{M/M_{\odot}}{(R/R_{\odot})^2 \sqrt{T_{eff}/T_{eff,\odot}}}
\end{equation}

provide a measure of the mass and radius. These scaling relations are very general, model independent, and an extremely useful tool for determining stellar parameters. It is, however, important that they be verified. Interferometric measurements of stellar radii have helped to confirm the validity of the scaling in radius \citep[see][for example]{huber2012}, but there have been limited tests of the seismic mass scale \citep[but see][]{miglio2012}. Because of the nature of the sharp transition between slow and rapid rotation as a function of mass, even on the subgiant branch, measurements of rotation period could provide a means to verify stellar masses derived via the scaling relations.

\subsubsection{Rotation as a Test of Stellar Physics}

We have presented here a set of very simple physical models: we consider only solid body rotation and AM losses from winds, both of which are well motivated by observations. Our previous discussion of the uses of rotation as a tool relies on the assumption that our models are correct. While our simple case is well motivated, there are more complicated effects, such as differential rotation, that could also be important, and we may find that our theoretical predictions do not agree with the data and motivate additional model complexity.

If we were to consider the addition of differential rotation, for example, we might consider 1) models that have internal differential rotation on the MS, and rapidly rotating cores on the SGB, 2) models that begin as solid bodies on the MS but develop internal differential rotation in radiative zones over the course of the SGB, or 3) models that allow for differentially rotating convective zones as well radiative zones. Each of these combinations would produce a different signature in the surface rotation rates of stars evolving on the SGB, and a pattern of departures from the simple, first-order case we have presented here. Case 1) would result in objects with faster rotation rates than we predict, as the base of the convective envelop dredges up material from deeper, more rapidly rotating regions of the star as the object evolves across the subgiant branch. In contrast, case 3), coupled with wind losses, could produce rotation periods longer than those of the solid body case, because surface layers depleted of AM would not be quickly resupplied with AM from the deeper interior.

Case 2) is among the simplest to address in a more qualitative way. We evolve a solid body model in YREC's fully rotating configuration of a 1.6 $M_{\odot}$ star to the MSTO. The model is then allowed to develop differential rotation as it undergoes core contraction and envelope expansion on the SGB. Radiative zones are assumed to conserve specific angular momentum, while convective zones always rotate as solid bodies. We neglect winds both on the MS and SGB for this exercise. Models of this form rotate a maximum of about 1.2 times more slowly (at $\sim 6100$ K) than a solid body model evolved under YREC's fully rotating configuration (which accounts for rotational deformation, unlike our standard models). This period difference is the result of a larger effective reservoir of AM available in the solid body case. There AM transport is assumed to be instantaneous, and therefore and AM that would otherwise be sequestered in the core is redistributed throughout the star immediately. In the case of the differentially rotating model, only the angular momentum of the outermost envelope is available, and thus we would observe a slower rotation rate. As the convective envelope deepens in both models the differences in the rotation rates decrease, because limited AM is located in the core of even the differentially rotating case, and both models have large envelopes rotating as solid bodies.

There are many other possibilities suitable for investigation that are beyond the scope of this paper. In the absence of large rotation datasets it has been impossible to determine which of these cases is most correct, and it is difficult to motivate the inclusion of additional model complexity. With enough observational data, it will be possible test more complex interiors models by recognizing the pattern of departures between the data and a simple set of models, such as those that we have provided in this paper. We have begun here with a very simple test case; as large datasets inform our theory, we may find that we are finally well motivated to include more complicated physics in our models. 

\subsection{In the Era of Gaia} \label{sec:gaia}

Measurements of rotation are highly complementary to the precise stellar parameters that will be obtained by the Gaia \footnotemark mission. Luminosities derived through parallaxes suffer from a mass-composition degeneracy, which means that although the object may be well-placed on a HR-diagram, fundamental parameters such as age and mass remain uncertain. Furthermore, luminosity in sensitive to helium, which unlike metal content, is not easily measured. As we have shown, rotation is also sensitive to composition, but its dependence is different: luminosity is set by the mean molecular weight, where rotation is set by the effective temperature and age of the star. The addition of rotation information can break the mass-composition degeneracy.

\footnotetext{http://sci.esa.int/science-e/www/area/index.cfm?fareaid=26}

Rotation also remains a means to identify unusual objects, such as stellar mergers and interacting systems, that fall in otherwise typical regions of the HR diagram but have unusual rotation periods \citep[e.g. subturnoff mergers, see][]{andronov2006}. Gaia will be able to identify likely binaries based on their HR diagram locations with respect to the single star MS, but rotation will be able to determine the fraction of those binaries that have synchronized. Massive binaries will appear as anomalously slowly rotating, whereas low mass binaries will have inflated rotation rates. Our predictions for typical rotation periods, coupled with observations and the precise HR diagram locations provided by Gaia, allow for interesting studies of binary evolution by isolating both pre- and post-merger objects.  

Photometric rotation information from Gaia itself will be limited \citep{distefano2012}, and the periods of solar analogs largely misidentified. However, \citet{distefano2012} suggests that for rapidly rotating objects with $P < 5$ days, the information from the satellite photometry alone may be adequate to identify the period, meaning that massive stars or synchronized binaries may have measured rotation periods. For broader populations, large-scale spot-modulation or rotation velocity surveys will be necessary to take full advantage of the constraining power of stellar rotation rates.

\section{Conclusions} \label{sec:conclusion}

Massive and precise datasets produced by spacecraft such as \textit{Kepler} and \textit{CoRoT} are transforming stellar and planetary astrophysics. Unlocking the full potential of these rich datasets requires the ability to sort through the complicated mixture of stars observed in field studies. As always, the challenge of sample characterization is to find an observable that varies strongly as a function of the parameters we wish to infer, and to exploit it. Rotation represents such an observable. Historically, it has been impractical to measure rotation in bulk field samples, but modern surveys will yield rotation periods for an unprecedented number of stars, making rotation a viable and useful tool. 

Rotation periods are shaped by stellar mass, age, and evolutionary state. A transition from rapid rotation at high mass to slower rotation at lower mass (the Kraft break) is imprinted on the MS and linked to the onset of AM loss from magnetized winds in the cool stars. Such winds progressively spin down cool stars as they age. As stars evolve, their moments of inertia also change, resulting in substantial changes in their surface rotation periods. Mass, age, and evolutionary state are therefore imprinted on the measured surface rotation rates.

We are already accustomed to using rotation as an age diagnostic in gyrochronology, but when we expand our studies beyond the realm of traditional gyrochronology and consider a mixed field population of both evolved and unevolved stars at many different masses, rotation can be used as a tool for a far broader set of investigations. It can provide constraints on stellar mass, evolutionary state, and radius in addition to its use as an age diagnostic, and becomes particularly useful on the subgiant branch, where these quantities can be otherwise difficult to obtain. These qualities make period measurements immediately useful for characterization of the hosts of transiting planets, where precision stellar radii are essential, and for the verification of the asteroseismic mass scale. 

We have shown that stellar rotation in a cool field population of MS and SGB stars falls into three distinct regimes:

\begin{enumerate}
\item{ $P < 10$ days: Young solar-like \textit{or} massive stars born above the Kraft break}
\item{ $10 < P < 40$ days: MS solar-like stars or crossing massive and intermediate mass subgiants.}
\item{ $P > 40$ days: Primarily solar/ low-mass subgiants}
\end{enumerate}

A star's period, in combination with an effective temperature, therefore immediately makes suggestions regarding its birth mass and current evolutionary state. These regimes also demonstrate that the presence of subgiants will be an important consideration when dealing with large, mixed stellar samples. Subgiants are numerous in magnitude-limited surveys, and represent an important background in the prime period range for gyrochronology (10-40 days). For effective temperatures in the range $5000-7000$ K, a survey with a magnitude limit of $K_p < 14$ will consist of roughly 35\% subgiants. The fraction in bright, seismically interesting samples ( $K_p < 11$) is even higher, at $\sim40\%$.  Any conclusions regarding the age distributions of a field sample determined through gyrochronology that do not account for presence of subgiants will be erroneous. It is essential that the contribution of subgiants to period distributions be recognized: they are paradoxically both an important contaminant and a population of stars for which stellar parameters are highly accessible through measurements of their rotation periods. 

We have presented here a set of predictions for the most simplistic physically motivated model of the evolution of the surface rotation rate, based on the existing rotation data. A sample of bright \textit{Kepler} stars with both measured rotation periods and seismic information will be available in the near future, and the exquisite precision of the data will provide powerful tests of these predictions that were not possible in the past. During the preparation of this manuscript, several large samples of rotation period have become available \citep{affer2012,nielsen2013}. Although beyond the scope of this paper, our next step is to compare these datasets with our theoretical predictions. The reality of stellar rotation may be far more complex than our simple model: it may be that stars leave the MS with substantial internal differential rotation. This would result in anomalously high observed rotation rates on the SGB as the convective envelope eats into the rapidly rotating interior of such a star. Likewise, it may be the case that the convective envelopes of these evolved stars support substantial differential rotation (as opposed to the theoretical expectation of solid body rotation). While the signature of this differential rotation would be present on the SGB as anomalously slow rotation, it could be confused with other effects, such as unexpectedly efficient winds, which could complicate the interpretation. We can however, examine the surface rotation rates giants after first dredge up; these objects would be sufficiently slowly rotating that winds should not be important, and the presence of differential rotation in the convective envelope could be more clearly identified. In this way, comparison of large datasets to our simple theoretical model will inform us as to whether or not additional complexity is needed to explain the observed distribution of rotation periods across the HR diagram, and we will then be able to motivate the inclusion of additional physics with observations. We stand to learn as much from disagreements between data and models as perfect agreement.

Once we have made these initial comparisons and refinements using small but excellent rotation datasets, rotation-based diagnostics will continue to be useful beyond the era of \textit{Kepler} and \textit{CoRoT}, and even in the presence of precise parallaxes (e.g. Gaia). Rotation depends on mass and composition in a fundamentally different manner than luminosity, which will help to break degeneracies between mass and metallicity that will exist even with exceptional parallax measurements. For large time-domain surveys in which parallaxes or extensive spectroscopic follow-up is unavailable, rotation remains a means to differentiate between dwarf and subgiant, evolved and unevolved, low-mass and high-mass stars.

\acknowledgements
We would like to thank Rafa Garc\'{i}a, Travis Metcalfe, Savita Mathur, and Benjamin Shappee for useful discussions and feedback. This research has made use of the WEBDA database, operated at the Institute for Astronomy of the University of Vienna. and was supported in part by the National Science Foundation of the United States under grant No. NSF PHY05-51164, the NSF Graduate Research Fellowship Grant RF\#743796 (J.V.S.), and the NASA grant NNX11AE04G (M.H.P.).

\bibliographystyle{apj}

\end{document}